\documentclass[a4paper, 12pt, oneside]{article}

\usepackage{aapreprint}

\usepackage{graphicx,wrapfig,float,slashed,subcaption,bbold,bm}
\usepackage{amsmath,amssymb,epsfig,graphicx,xcolor}
\usepackage{epstopdf}
\usepackage{booktabs}
\usepackage{ragged2e}
\usepackage{mciteplus}
\usepackage{mathtools}
\usepackage{adjustbox}
\allowdisplaybreaks

\usepackage{cancel}

\usepackage{makecell}

 \usepackage{comment}

\newcommand{\nc}{\newcommand}
\nc{\non}{\nonumber}
\nc{\hc}{\hbox {H.c.}}
\nc{\noi}{\noindent}
\nc{\barx}{\bar{x}}
\nc{\pbarn}{\;\hbox {pb}}
\nc{\fbarn}{\;\hbox {fb}}

\nc{\hsp}{\hspace{0.5cm}}
\nc{\lsp}{\hspace{1cm}}
\nc{\Lsp}{\hspace{2cm}}
\nc{\LLsp}{\lsp\lsp}
\nc{\lra}{\longrightarrow}
\nc{\p}{\prime}
\nc{\sgn}{\text{sgn}}
\nc{\ph}{\varphi}
\nc{\op}{{\cal O}}

\nc{\beq}{\begin{equation}}  \nc{\eeq}{\end{equation}}
\nc{\bea}{\begin{eqnarray}}  \nc{\eea}{\end{eqnarray}}
\nc{\baa}{\begin{array}}     \nc{\eaa}{\end{array}}
\nc{\bit}{\begin{itemize}}   \nc{\eit}{\end{itemize}}
\nc{\ben}{\begin{enumerate}} \nc{\een}{\end{enumerate}}
\nc{\bce}{\begin{center}}    \nc{\ece}{\end{center}}
\nc{\bpm}{\begin{pmatrix}}   \nc{\epm}{\end{pmatrix}}
\nc{\bvt}{\begin{verbatim}}  \nc{\evt}{\end{verbatim}}

\def\lsim{\mathrel{\raise.3ex\hbox{$<$\kern-.75em\lower1ex\hbox{$\sim$}}}}
\def\gsim{\mathrel{\raise.3ex\hbox{$>$\kern-.75em\lower1ex\hbox{$\sim$}}}}

\def\udots{\mathinner{\mkern1mu\raise1pt\vbox{\kern7pt\hbox{.}}\mkern2mu\raise4pt\hbox{.}\mkern2mu\raise7pt\hbox{.}\mkern1mu}}

\def\dd{\mathrm d}

\newcommand{\Eq}[1]{Eq.~(\ref{#1})}

\newcommand\fverb{\setbox\fverbbox=\hbox\bgroup\verb}
\newcommand\fverbdo{\egroup\medskip\noindent%
			\fbox{\unhbox\fverbbox}\ }
\newcommand\fverbit{\egroup\item[\fbox{\unhbox\fverbbox}]}
\newbox\fverbbox

\preprint{\begin{flushright}
\end{flushright}}

\title{Spontaneous CP Breaking in  a QCD-like Theory} 
\author[a]{Csaba Cs\'aki,}
\author[a]{Maximilian Ruhdorfer,}
\author[a,b]{and Taewook Youn}
\affiliation[a]{Laboratory for Elementary Particle Physics \\ Cornell University, Ithaca, NY 14853, USA}
\affiliation[b]{Department of Physics, Korea University \\ Seoul 02841, Republic of Korea}
\emailAdd{csaki@cornell.edu}
\emailAdd{m.ruhdorfer@cornell.edu}
\emailAdd{taewook.youn@cornell.edu}

\abstract{We examine the phase structure of a QCD-like theory at $\bar\theta =\pi$ obtained from supersymmetric $SU(N)$ QCD perturbed by a small amount of supersymmetry breaking via anomaly mediation (AMSB QCD). The spectrum of this theory matches that of QCD at the massless level, though the superpartners are not decoupled. In this theory it is possible to nail down the phase structure at $\bar\theta=\pi$ as a function of the  quark masses and the number of flavors $F$. For one flavor we find that there is a critical quark mass, below which CP is unbroken, while above the critical mass CP is spontaneously broken. At the critical mass there is a second-order phase transition along with a massless $\eta'$. We are able to analytically solve for the minima and the critical mass for $N=2,3$ as well as for the large $N$ limit, while for other $N$ one can find numerical results. For two flavors, we find that CP is always broken as long as the quark masses are equal and non-zero, however there is a non-trivial phase boundary for unequal quark masses, which we find numerically. For $F\geq 3$ we obtain an intricate phase boundary which reproduces the various quark mass limits. All our results are in agreement with the predictions of Refs.~\cite{Gaiotto:2017tne,Gaiotto:2017yup,DiVecchia:2017xpu,Cordova:2019bsd,Cordova:2019jnf} for ordinary QCD that were based on anomaly matching arguments for generalized symmetries and the effective chiral Lagrangian. 
We also briefly comment on the domain wall solutions first discussed by Draper~\cite{Draper:2018mpj}, and are able to present analytic results for the simplest case of $SU(2)$ with one flavor.   }

\begin{document}

\maketitle
\flushbottom

\section{Introduction}
It is well known that $\theta$, the topological angle, plays an important role in the vacuum structure of QCD or QCD-like theories. For generic $\theta$, time-reversal symmetry, or equivalently CP, is broken and QCD is believed to be gapped with a unique vacuum. However, for $\theta=\pi$ the story is more complicated. While CP is not explicitly broken for this value it appears to be spontaneously broken for a wide range of choices for $N,F$ and quark masses. Let us make this statement more precise in the following.

At large $N$ it can be made rigorous that $SU(N)$ Yang-Mills theory has a unique vacuum for generic $\theta$ but two degenerate vacua at $\theta =\pi$ such that CP is spontaneously broken~\cite{Witten:1978bc,Witten:1979vv,Witten:1980sp,DiVecchia:1980yfw,Nath:1979ik,Witten:1998uka,Kawarabayashi:1980dp,Ohta:1981ai}. Furthermore when $\theta$ is moved through $\theta = \pi$ there is a first-order phase transition. This picture can be extended to finite $N$ using anomaly arguments. For pure $SU(N)$ Yang-Mills theory there is a mixed 't~Hooft anomaly between its $\mathbb{Z}_N$ one-form center symmetry and time-reversal symmetry which can be used to argue that the phase transition still exists at finite $N$~\cite{Gaiotto:2017yup}.

When quarks in the fundamental representation are introduced the picture becomes more complicated. The theory does not possess a $\mathbb{Z}_N$ one-form center symmetry anymore, such that the anomaly argument for pure $SU(N)$ Yang-Mills theory does not apply. For massless quarks $\theta$ is unphysical and has no effect on the vacuum structure. Once masses are turned on, the theory depends on the effective topological angle $\bar{\theta} = \theta + \arg\det M_Q$, where $M_Q$ is the quark mass matrix. At large $N$ it can be argued that for $F>1$ flavors with equal masses $M_Q = m_Q \mathbb{1}_{F\times F}$, CP is spontaneously broken for any $|m_Q | > 0$ at $\bar{\theta} = \pi$ with a first order phase transition when $\bar{\theta}$ moves through $\bar{\theta}=\pi$~\cite{Gaiotto:2017tne,DiVecchia:2017xpu}. This is consistent with anomaly arguments that use that the theory with $SU(F)$ flavor symmetry, i.e. equal masses, has a mixed 't~Hooft anomaly of a global $U(F)/\mathbb{Z}_N$ symmetry\footnote{Naively the global symmetry appears to be $SU(F)\times U(1)$, where $U(1)$ can be identified with the baryon number. However, only $U(F)=(SU(F)\times U(1))/\mathbb{Z}_F$ acts faithfully on the fundamental representation. Additionally a $\mathbb{Z}_N$ subgroup has the same effect as gauge transformations in the center of $SU(N)$ and therefore is a gauge symmetry. Thus the global symmetry which acts faithfully is $U(F)/\mathbb{Z}_F$.} with time-reversal symmetry for some values of $N$ and $F$~\cite{Cordova:2019jnf,Cordova:2019uob}. However for $F=1$ or $F>1$ with non-degenerate quark masses there is no remaining symmetry for such an argument. And indeed at large $N$ one finds that there is a critical mass value for $F=1$ and a critical hypersurface for $F>1$ in the space of masses which separates a region where CP is spontaneously broken from a region where it is unbroken~\cite{Gaiotto:2017tne,DiVecchia:2017xpu}. The phase transition is second order with massless degrees of freedom at the critical point or on the critical hypersurface.

At finite $N$ the vacuum structure can be studied using the phenomenological chiral Lagrangian. In $F=3$ flavor QCD one finds a region of quark masses for which CP is spontaneously broken~\cite{Smilga:1998dh,Creutz:2003xu,Creutz:2006ts}. However, for $F=2$ higher order terms in the chiral Lagrangian, which cannot be computed from first principles, are required to assess if CP is spontaneously broken~\cite{Smilga:1998dh}. And for $F=1$ there is only one would-be Goldstone boson whose shift-symmetry is badly broken by the chiral anomaly. Studying such special cases cannot be done within the chiral Lagrangian framework alone and requires an extended theoretical toolbox.

\begin{figure}[t]
\centering
    \includegraphics[width=0.8\textwidth]{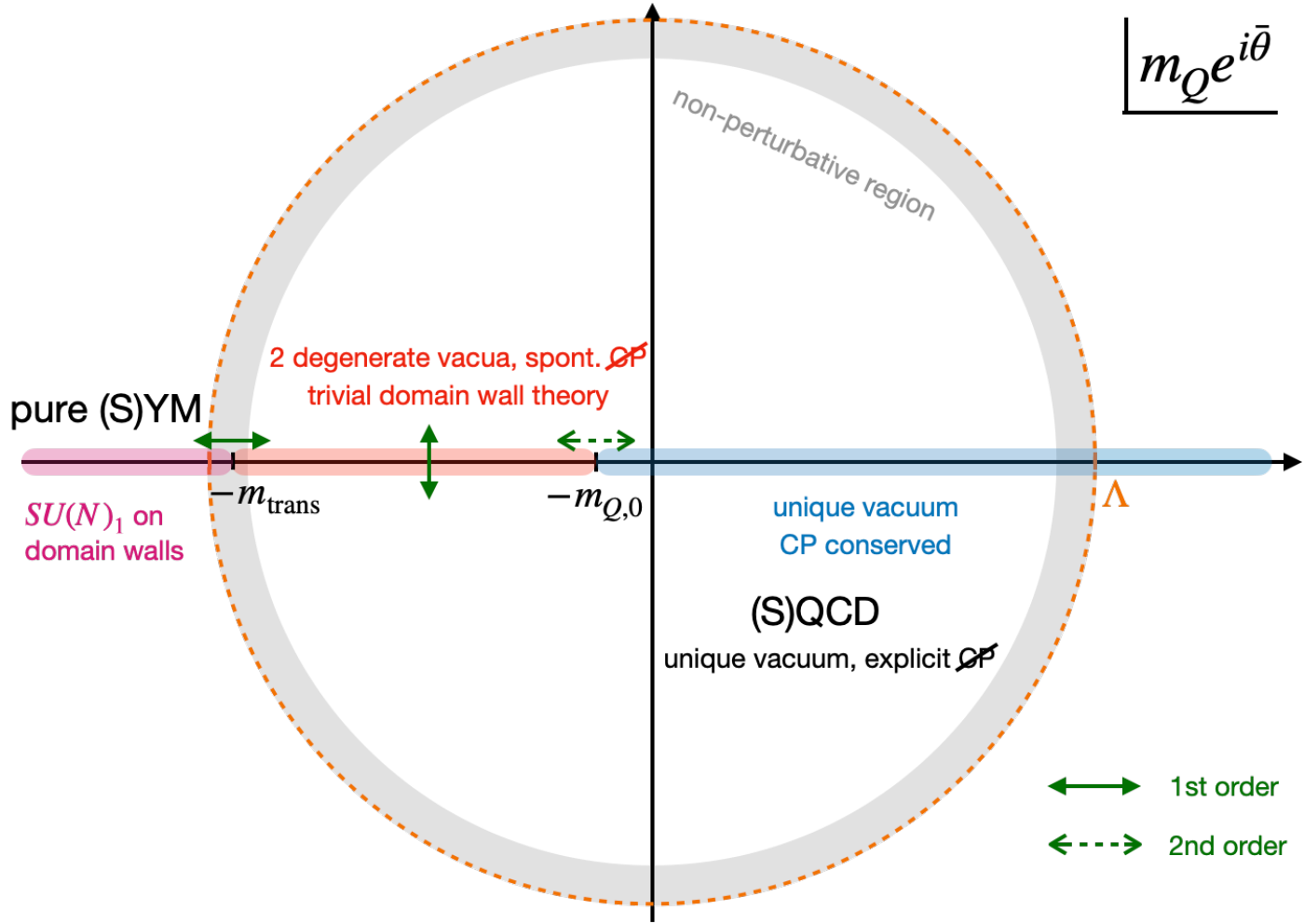}
\caption{Schematic overview of the phase structure of $F=1$ (S)QCD in the complex mass plane $m_Q e^{i\bar{\theta}}$ with $m_Q > 0$. The positive real axis corresponds to $\bar{\theta} = 0$ where CP is conserved (blue), whereas the negative real axis depicts the $\bar{\theta} = \pi$ direction where CP is spontaneously broken above a critical value $m_{Q,0}$ (red and purple). For $m_Q > \Lambda$ the low-energy theory reduces to a pure (S)YM theory. Green arrows show the order of the phase transition between different phases when parameters are varied along the indicated direction.}
\label{fig:phaseStructure}
\end{figure}

Ref.~\cite{Gaiotto:2017tne} recently presented a coherent QCD phase structure based on the above large $N$ arguments that is also consistent with anomaly arguments and the expectations for the $m_Q \gg \Lambda$ and $m_Q = 0$ limits. For $|m_Q | \gg \Lambda$ one recovers pure Yang-Mills theory where CP is spontaneously broken at $\bar{\theta} =\pi$, while for $m_Q=0$ the topological angle $\bar{\theta}$ is unphysical and we obtain CP-conserving vacua. Thus for non-degenerate quark masses (for degenerate masses see the discussion above) there has to be a transition, i.e. a critical point or a critical hypersurface in the quark mass parameter space, at which the theory with a unique CP conserving vacuum transitions into a theory with spontaneously broken CP. This situation is schematically shown in Figure~\ref{fig:phaseStructure} for a single light flavor. At energies $E\ll \Lambda$ the theory depends only on one complex parameter $m_Q e^{i\bar{\theta}}$ where we can take $m_Q > 0$ without loss of generality. For generic ${\bar\theta} \neq \pi$ there is a unique vacuum and CP is explicitly broken for ${\bar \theta} >0$. Along the positive real axis, i.e. for $\bar{\theta}= 0$ CP is always conserved, while for $\bar{\theta} = \pi$, i.e. along the negative real axis, there is a critical value $m_{Q,0} \sim \Lambda / N$~\cite{Gaiotto:2017tne} above which CP is spontaneously broken. Note that the transition can only be reliably studied in the large $N$ limit when $m_{Q,0} \ll \Lambda$ such that the theory has a weakly-coupled description in terms of light Goldstone bosons. The transition between these phases along the real axis is second order with a massless degree of freedom at the critical point, while the transition along the ${\bar\theta}$ direction is first order.  
When CP is spontaneously broken there are dynamical domain walls. In the case that $m_{Q,0} < m_Q \ll \Lambda$ the theory on the domain wall is trivial for $F=1$ and accommodates a $\mathbb{CP}^{F-1}$ non-linear sigma model with a Wess-Zumino term for $F>1$. When $m_Q \gtrsim \Lambda$ the low-energy theory is a pure Yang-Mills theory which has a $SU(N)_{1}$ Chern-Simons theory on the domain wall. Thus there has to be a critical value $m_{\rm trans}$ at which a phase transition occurs on the domain wall (see e.g.~\cite{Gaiotto:2017tne}). This occurs when the low-energy theory becomes strongly coupled.

While some parts of this phase diagram can be studied using the chiral Lagrangian, especially cases with a small number of flavors (e.g. $F=1,2$) are only accessible at large-$N$. For this reason it would be beneficial to study these features in a QCD-like theory which allows to reliably determine the phase structure for a wider range of $F$ and $N$ values. This is where softly broken supersymmetric QCD (SQCD) has its strengths. It allows us to study these features at finite and even small $N$ and $F$ in a theory that is under control but has many of the properties of QCD, such as confinement and chiral symmetry breaking. 

\subsection{CP Breaking in SQCD: Results}\label{subsec:CPinSQCD}
%
Softly broken SQCD allows us to study the full phase diagram of the low-energy theory even for a small number of flavors, such as $F=1,2$, which are inaccessible in chiral perturbation theory. The reason for this is twofold: i) the chiral Lagrangian in softly broken SQCD can be derived from a top-down perspective (see e.g.~\cite{Dine:2016sgq,Csaki:2023yas}) such that higher-order terms are known in terms of the fundamental parameters, and ii) the presence of a spontaneously broken non-anomalous combination of the $U(1)_R$ and $U(1)_A$ symmetry, which only gets explicitly broken by the quark masses and soft SUSY breaking. This results in a light $\eta'$ in SQCD, i.e. $m_{\eta'} \sim m \ll \Lambda$, in contrast to ordinary QCD. As a result, the critical value found for $F=1$ scales as $m_{Q,0}\sim m/N \ll \Lambda$, where $m$ is the SUSY breaking scale, whereas $m_{Q,0} \sim \Lambda/N$ in ordinary QCD. 

We find that due to the lightness of the $\eta'$ the phase structure closely resembles that of large-$N$ QCD which was outlined in~\cite{DiVecchia:2017xpu}. However, our results also extend to small $N$. For $F=1$ and $N=2,3$ we find an analytically calculable critical point below which CP remains unbroken at $\bar\theta = \pi$ and above which it is spontaneously broken. This picture extends to a larger number of flavors where the critical point is extended to a critical hypersurface. This hypersurface can be found numerically. Our analytic solutions for the minima also allow us to find both the domain wall profile and tension in the $F=1$ case.

The paper is organized as follows. We start in Section~\ref{sec:Vacuum} with a short reminder about SQCD with AMSB and discuss the vacuum structure in the limiting cases of vanishing SUSY-breaking $m\rightarrow 0$ and vanishing supersymmetric masses $m_Q\rightarrow 0$. In Section~\ref{sec:CPPhaseStructure} we present our results on the CP phase structure for various choices of $F$ and $N$. We furthermore comment on domain wall solutions which connect the degenerate CP breaking vacua in Section~\ref{sec:DomainWall}. Finally we conclude in Section~\ref{sec:conclusion}.

\section{Reminder about AMSB SQCD}\label{sec:Vacuum}
%
The dynamics and vacuum structure of SQCD crucially depend on the number of colors $N$ and the number of flavors $F$. In the simplest case where $F<N$, which we will mainly focus on in the following, the low-energy degrees of freedom are parameterized in terms of the meson superfield $M_{f f'} = \bar{Q}_f Q_{f'}$, where $f,f'$ are flavor indices and $Q_f, \bar{Q}_f$ are the quark chiral superfields. The low-energy dynamics is determined by the non-perturbative ADS superpotential~\cite{Davis:1983mz,Affleck:1983mk}
\beq \label{eq:NgtrFW}
W = (N - F) \left(\frac{\Lambda^{3N-F}}{\det M} \right)^{\frac{1}{N-F}} + \text{Tr}(m_Q M)
\,.
\eeq
where we additionally included a supersymmetric mass term for the quarks. Note that using $SU(F)_L \times SU(F)_R$ flavor transformations, $m_Q$ can always be brought to the form $m_Q = e^{i\theta_Q} \text{diag} (m_{Q,1},\ldots , m_{Q,F})$. 

We additionally assume that $\mathcal{N}=1$ SUSY is softly broken via anomaly mediation (AMSB)~\cite{Randall:1998uk,Giudice:1998xp,Arkani-Hamed:1998mzz,Luty:1999qc,Pomarol:1999ie}. Denoting the SUSY breaking scale by $m$ one finds a scalar potential, which at tree-level is given by~\cite{Murayama:2021xfj,Csaki:2022cyg}
\beq
V_\mathrm{tree} = \partial_i W g^{ij^*} \partial^*_j W^* + m^* m \bigg( \partial_i K g^{ij^*} \partial^*_j K - K \bigg) + m \bigg( \partial_i W g^{ij^*} \partial^*_j K - 3W \bigg) + {\rm h.c.},
\label{eq:vtree}
\eeq
where $K$ is the K\"ahler potential, $ g_{i j^*} = \partial_i \partial_{j}^\ast K$ is the K\"ahler metric, and $g^{ij^*}$ is its inverse. The SUSY breaking scale is in general complex, i.e. $m = e^{i\theta_m} |m|$. However, $m$ explicitly breaks $U(1)_R$ and $m_Q$ breaks $U(1)_A \times U(1)_R$, such that a combination of $U(1)_R$ and $U(1)_A$ transformations can be used to rotate $\theta_Q$ and $\theta_m$ into $\theta$, the vacuum angle of the gauge theory. Thus the theory contains only one physical phase which we can express in terms of the $U(1)_A \times U(1)_R$ invariant combination\footnote{This can be easily seen by promoting $m, m_Q$ and $\Lambda$ to spurions which under $(U(1)_A,U(1)_R)$ transform as $\Lambda^{3N-F}\sim (2F,0),\, m_Q \sim (-2, 2N/F)$ and $m\sim (0,-2)$. This fixes the $U(1)_A\times U(1)_R$ invariant phase to be $\arg (m_Q^F \Lambda^{3N-F} m^N ) =\theta + F\theta_Q + N \theta_m$.}
\beq
\bar{\theta} = \theta + F \theta_Q + N \theta_m\,.
\eeq
Replacing $\theta$ in the holomorphic scale $\Lambda$ by $\bar{\theta}$ we can take without loss of generality all entries of $m_Q$ to be real and $m > 0$. 
%
\subsection{Moduli Space and Vacuum Structure} \label{subsec:vacuum}
%
The classical moduli space, parameterized by the meson superfield $M$, is lifted by the non-perturbative ADS superpotential. However, the scalar potential originating from the ADS superpotential has a runaway direction, such that the theory does not possess a ground state. Stabilizing the potential requires the addition of a supersymmetric quark mass term as in Eq.~\eqref{eq:NgtrFW} or soft SUSY breaking. In general both are present and their interplay leads to a nontrivial vacuum structure of SQCD when they are roughly of the same size. However, it is nonetheless instructive to consider both limiting cases $m_Q \rightarrow 0$ and $m\rightarrow 0$ as they will give us an intuition about the asymptotics in the general case.
%
\subsubsection{Potential Stabilization by SUSY Quark Masses}\label{sec:PotMassStabilization}
%
In the supersymmetric limit $m\rightarrow 0$ the vacuum is determined by the critical point of the superpotential, i.e. $d W / (d M_{ff'}) =0$, which is solved by
\beq \label{eq:SUSYVacuaMass}
\begin{split}
\langle M_{f f'}\rangle 
=  (m_Q^{-1})_{f f'} \left( \det (m_Q) |\Lambda|^{3N-F}\right)^{1/N} e^{i\frac{\bar{\theta} + 2\pi k}{N}}\,,\quad k=0,\ldots , N-1\,,
\end{split}
\eeq
i.e. there are $N$ degenerate vacua, originating from the complex $N$-th root of $\Lambda$. Note that at a generic point in moduli space the meson VEVs break the gauge group to $SU(N-F)$ which gets strong in the IR and confines. The low-energy superpotential of the pure $SU(N-F)$ gauge theory is multi-valued with $N-F$ branches, corresponding to the $N-F$ branches of the gaugino condensate\footnote{Note that the ADS superpotential can be interpreted as being generated by gaugino condensation in the unbroken $SU(N-F)$ group, i.e. $W_{\rm eff} = (N-F) \left( \Lambda^{3N-F}/ \det M\right)^{1/(N-F)}$.}
\beq
\left\langle \lambda^a \lambda^a \right\rangle = 16\pi i \frac{\partial W_{\rm eff}}{\partial \tau} = -32\pi^2 \left(\frac{\Lambda^{3N-F}}{\det M} \right)^{\frac{1}{N-F}}\,.
\eeq
If we restrict $M$ to the form $M = |M| e^{i\eta'}$ the phase of the gaugino condensate is given by
\beq
\arg \left\langle \lambda^a \lambda^a \right\rangle = \frac{\bar{\theta} - F\, \eta' + 2\pi j}{N-F}\,,\quad j=0,\ldots,N-F-1\,.
\eeq
For a fixed $\eta'$ there are $N-F$ different branches of the gaugino condensate and a $2\pi$ rotation of $\eta'$ connects branch $j$ with branch $j-F$~mod~$(N-F)$. Thus, there are gcd$(F,N-F)=$gcd$(F,N)$ components which cannot be continuously connected through phase rotations in the $\eta'$ direction alone (see e.g.~\cite{Draper:2018mpj,Bashmakov:2018ghn} for further explanations). The vacua in Eq.~\eqref{eq:SUSYVacuaMass} are distributed over the $N-F$ branches and can be continuously connected through the $\eta'$ direction if gcd$(F,N)=1$. If gcd$(F,N)>1$ some vacua have to be connected through directions that do not only have a global phase.

Adding a small amount of SUSY breaking via anomaly mediation, i.e. we assume $m\ll m_Q$ such that the potential is still stabilized by the quark masses, the degeneracy of the vacua is broken
\beq
\begin{split}
    \delta V_{\rm AMSB} &= -m\left[ (3N - F) \left(\frac{\Lambda^{3N-F}}{\det M} \right)^{\frac{1}{N-F}} + \text{Tr}(m_Q M) \right] + h.c.\\
    &= -6N m\, (\det m_Q)^{1/N} |\Lambda |^{(3N-F)/N} \cos\left(\frac{\bar{\theta} + 2\pi k}{N}\right)\,,
\end{split}
\eeq
which for general $\bar{\theta}$ has a single value of $k$ which minimizes the potential. For the corresponding expression for general soft masses for the gauginos and squarks see~\cite{Draper:2018mpj}. Note however that in such a scenario there are two invariant phases due to different phases in the soft gaugino and squark masses. 

In the above expression there are two values of $\bar{\theta}$ which are of special interest: $\bar{\theta} = 0$ for which the minimum occurs at $k=0$ and $\bar{\theta} =\pi$ which has two degenerate minima at $k=0$ and $k=N-1$. CP acts on the mesons as $M\xrightarrow{\rm CP} M^\dagger$. This implies that the single minimum at $\bar{\theta} =0$ for $k=0$ is as expected CP conserving. However, the two minima at $\bar{\theta} =\pi$ are CP conjugates and therefore CP is spontaneously broken. In such a scenario there exist domain walls which interpolate between these degenerate vacua. Depending on $F$ and $N$ these vacua might be on the same branch of the $SU(N-F)$ gaugino condensate which would allow them to be continuously connected by phase rotations of $\det M$ (i.e. in the $\eta'$ direction which is the would-be Goldstone boson of the spontaneously broken $U(1)_A$) or they might be on different branches which requires a deformation along directions of $M$ which break the residual flavor symmetry. A thorough discussion of both types of domain walls can be found in~\cite{Draper:2018mpj}. Here we are more interested in the transition from the $m_Q \gg m$ region where CP is always spontaneously broken at $\bar{\theta} =\pi$ to the opposite region $m\gg m_Q$ where this is not the case.
%
\subsubsection{Potential Stabilization by SUSY Breaking}\label{subsec:SUSYBreakingStabilization}
%
Let us now discuss the opposite scenario where the scalar potential is stabilized by SUSY breaking effects, i.e. we take the limit $m_Q \rightarrow 0$. In this case the theory possesses an exact $SU(F)_L \times SU(F)_R$ flavor symmetry and we can make the ansatz $\langle M_{f f'} \rangle = f^2 U_{f f'}$, where $U\in SU(F)$, for the minimum. In terms of $f$ the scalar potential including AMSB effects takes the form
\beq
V = (2F)^{-1} \left| \frac{2F}{f}\left(\frac{\Lambda^{3N-F}}{f^{2F}}\right)^{1/(N-F)} \right|^2 - m(3N-F)\left(\frac{\Lambda^{3N-F}}{f^{2F}}\right)^{1/(N-F)} + h.c.
\eeq
The minima can be parameterized as
\beq \label{eq:mQllmVEV}
M = |\Lambda|^2 \left(\frac{N+F}{3N-F} \frac{|\Lambda |}{m}\right)^{(N-F)/N} e^{i\frac{\bar{\theta}}{F}}\, U\,,
\eeq
where $U$ is an arbitrary $SU(F)$ matrix. A few comments are in order. As long as $m_Q = 0$ the above meson VEV eliminates the dependence on $\bar{\theta}$ from the potential making it manifest that $\bar{\theta}$ is unphysical in the presence of massless fermions and $\eta'$ acts as a QCD axion. We should also note that for $F=1$ one should make the replacement $U\rightarrow 1$ which implies that there exists only one unique minimum in this limit which 
leads to a real meson VEV for $\bar{\theta}=0,\pi$, implying that CP remains always unbroken in this limit. This is in contrast to the situation in the previous subsection where we found that for $m_Q \gg m$ CP is always spontaneously broken at $\bar\theta = \pi$. This conclusion remains true even when a small mass $m_Q\ll m$ is introduced and implies that there must be a transition from the unbroken phase to the broken phase when $m_Q$ is increased. In the next section we will find that this transition occurs at $m_{Q,0} \sim m/N$. 

For $F>1$ the introduction of masses lifts the degeneracy of the vacua. Parameterizing $U$ as
\beq \label{eq:Uparam}
U = \text{diag} (e^{i \alpha_1}, e^{i\alpha_2},\ldots, e^{i\alpha_F})\,, \quad \text{with}\quad 0\leq \alpha_i < 2\pi\,,\quad \sum_{i=1}^F \alpha_i = 0\quad \text{mod }2\pi\,,
\eeq
we find to linear order in $m_Q$ the following correction to the potential
\beq 
\label{eq:smallMassPot}
\delta V_{m_Q} \propto - |\Lambda |^3 \left(\frac{m}{|\Lambda|}\right)^{F/N} \sum_{i=1}^F m_{Q,i}\, \cos\left(\frac{\bar{\theta}}{F} + \alpha_i\right)\,,
\eeq
where the proportionality constant is a positive function of $F$ and $N$. For generic $\bar{\theta}$ there is a single minimum, whereas for $\bar{\theta} = \pi$ the vacuum structure depends crucially on the mass hierarchy. For equal quark masses $m_{Q,1} = \ldots = m_{Q,F} \equiv m_Q$ there is a $SU(F)_V$ flavor symmetry that remains unbroken after the introduction of the mass term. Whether this flavor symmetry remains unbroken will be determined by the meson VEVs. For $F>2$ we find that Eq.~\eqref{eq:smallMassPot} is minimized by
\beq
\alpha_1 =\ldots = \alpha_F = \frac{2\pi k}{F} \,,\quad \text{with} \quad k=0, F-1\,.
\eeq
Thus there are two degenerate minima for $U$ that lead to complex meson VEVs. These VEVs lie within the center of $SU(F)$ and therefore leave the $SU(F)_V$ flavor symmetry unbroken but they break CP spontaneously. Note that $F=2$ is special since Eq.~\eqref{eq:smallMassPot}, i.e. the potential to leading order in $m_Q$, vanishes. Finding the VEV requires the inclusion of higher orders in $m_Q$. While this corresponds to higher-order operators in chiral perturbation theory in QCD, for which there is no theoretical prediction (see e.g.~\cite{Smilga:1998dh}), in softly broken SQCD higher-order contributions to the potential can be computed. We will present numeric and approximate analytic solutions for the $F=2$ case in Section~\ref{subsec:NlargerF2}.

Away from the equal-mass limit the vacuum structure depends on the mass hierarchy. If one of the masses is significantly larger than the others it becomes favorable to relax the argument of the corresponding cosine in Eq.~\eqref{eq:smallMassPot} to zero, e.g. if $m_{Q,1} \gg m_{Q,2},\ldots, m_{Q,F}$ one finds $\alpha_1 = (2F -1)\pi/F$. In such a case the heaviest mass eigenstate is essentially integrated out and one recovers the potential for $F-1$ flavors. If such a hierarchy is present for all masses, i.e. $m_{Q,1} \gg m_{Q,2}\gg \ldots \gg m_{Q,F}$ there will be a unique CP-conserving vacuum with $\alpha_1 = \ldots = \alpha_{F-1} = (2F -1)\pi/F$ and $\alpha_F = (F-1)\pi/F$ where the arguments of the $F-1$ cosines with the largest masses are relaxed to zero, while $\alpha_F$ is determined by the requirement that $U\in SU(F)$. The exact boundary between these phases can only be found numerically in general.

Before moving on, let us note that the $m_Q \ll m$ limit is closer to real world QCD than the opposite limit. The SUSY breaking scale $m$ plays the role of the QCD scale since it sets the mass scale of the $\eta'$ boson which in the limit $m_Q \ll m$ is heavy and can be integrated out. Its VEV is responsible for the $e^{i\bar{\theta}/F}$ factor in the meson VEV in Eq.~\eqref{eq:mQllmVEV}. The vacuum structure is consequently set by the light pion-like directions in the special unitary matrix $U$, in complete analogy with ordinary chiral perturbation theory. If  $m_Q \sim m$ instead, the $\eta'$ is light and has to be included in the potential minimization. This scenario is conceptually similar to large $N$ QCD, where the $\eta'$ mass is suppressed by $1/N$. Note, however, that the lightness of the $\eta'$ in SQCD also holds for finite $N$.

\section{CP Phase Structure}\label{sec:CPPhaseStructure}
%
We are now ready to investigate the detailed phase structure of SUSY QCD with AMSB at  $\bar\theta = \pi$. 
First we will be discussing the case of a single flavor $F=1$, where a critical quark mass is expected to correspond to a second-order phase transition (PT) between the broken and unbroken CP phases. For a small number of colors ($N=2,3$) we will be able to find analytic results, as well as for the large $N$ limit, while for $N>3$ we can tackle the problem numerically. We will then discuss the case of a higher number of flavors. In agreement with the results from considerations in ordinary QCD we find that for equal quark masses CP is always broken as long as the quark masses are non-vanishing. However, once the quark masses are not equal, there will be a non-trivial phase boundary, as suggested by the limiting case when one of the quark masses is much heavier than the other. 

In order to investigate the phase structure at $\bar\theta = \pi$, we stabilize the potential obtained from \Eq{eq:NgtrFW} and \Eq{eq:vtree} by the most general spectra of SUSY quark masses $m_Q$ and SUSY breaking $m$. 
Subsequently, we find the global minimum, parameterized by the meson VEV $\langle M_{f f'}\rangle$, for $\bar\theta = \pi$. Time-reversal symmetry is spontaneously broken when there are degenerate minima which are related by the CP transformation $M\xrightarrow{\rm CP} M^\dagger$.
Note that while we always need to stick to the limit $m\ll \Lambda$ in order to have reliable results close to the SUSY limit, the quark mass $m_Q$ is holomorphic, so we are able to take both the $m_Q\gg \Lambda$ and $m_Q\to 0$ limits.  

\subsection{$F=1$: Exact Results}\label{subsec:F1Results}
We start with the $F=1$ case (and obviously $F < N$), where the phase structure can be studied analytically for a few small numbers of colors ($N=2, 3$) and in the large $N$ limit, as we will show below. For a detailed derivation of the results presented here, see App.~\ref{sec:potandmin}. 

For $F=1$ SQCD, the meson $M = \phi^2$ can be written in terms of a single chiral superfield $\phi$ with canonical K\"ahler potential $K=2\, \phi^\dagger \phi$.~\footnote{For $N=2$, the global flavor symmetry is actually enhanced $SU(2)$. However, this $SU(2)$ is not broken along the $D$-flat direction, hence it does not produce additional Goldstone bosons, while an anomaly-free combination of $R$ and global $U(1)$ symmetries is broken.} The resulting scalar potential can be found from~\Eq{eq:vtree} and is given by
\begin{equation}
V = \frac{1}{2} \left| \frac{2}{\phi} \left( \frac{\Lambda^{3N - 1}}{\phi^{2}}\right)^{\frac{1}{N-1}} - 2 \phi\, m_Q \right|^2 - m\left[(3N-1)  \left( \frac{\Lambda^{3N - 1}}{\phi^{2}}\right)^{\frac{1}{N-1}} + \phi^2 m_Q   \right] + {\rm h.c.}\,.
\label{eq:FsmallerNPot}
\end{equation}
Separating the magnitude and phase of the complex scalar $\phi$, i.e. $\phi = f \exp(i\delta_f)$, and introducing the variable
\begin{equation}\label{eq:xVar}
x = |\Lambda| \left( \frac{|\Lambda|}{f}\right)^{\frac{2N}{N-1}} 
\end{equation}
the potential can be written as $V = \min_\ell V_\ell$ with
\begin{equation}
\begin{split}
	V_\ell = 2\, x^{\frac{1-N}{N}} |\Lambda|^{\frac{3N-1}{N}} &\left(m_Q^2 + x^2 - m m_Q \cos (2 \delta_f) -   m (3 N -1) x \cos\left[\frac{2 \delta_f -  \bar\theta - 2\pi \ell}{N-1}\right] \right.\\
	&\left. \qquad - 2 m_Q x \cos\left[\frac{
     2 N \delta_f -  \bar\theta - 2\pi \ell}{N-1}\right]\right)\,,\quad \ell = 0,1,\ldots,N-2\,,
     \end{split}
     \label{eq:vlf1}
\end{equation}
where $2\pi \ell / (N-1)$ arises from the complex root and the explicit dependence on $| \Lambda |$ is factored out thanks to the variable transformation in Eq.~\eqref{eq:xVar}. Additionally we can restrict the SUSY quark mass to $m_Q \geq 0$ since we can absorb the sign into the definition of $\bar\theta$.
Note that the potential has no branches for $SU(2)$ ($\ell = 0$) and starts to be branched from $SU(3)$ ($\ell = 0,1)$ onward. The physical potential is given by the branch with minimal energy at each point, i.e. the lower envelope of all branches. 

\begin{figure}[t]
\centering
  \includegraphics[width=0.46\textwidth]{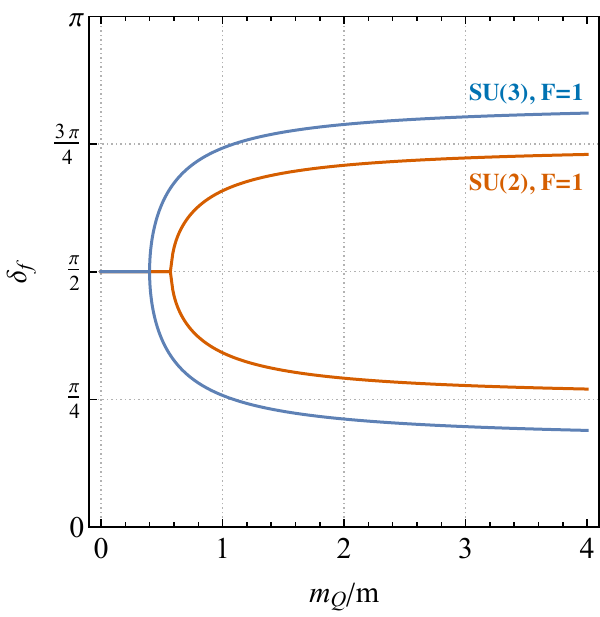}
  \hfill
    \includegraphics[width=0.498\textwidth]{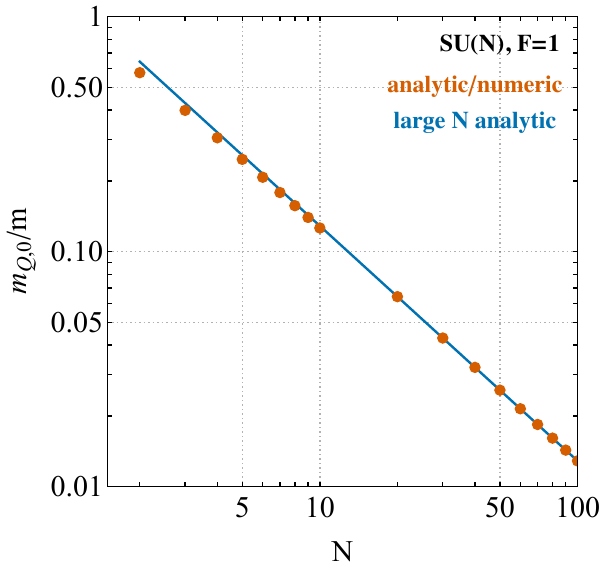}
\caption{Left: Phase $\delta_f$ at the minimum of the potential as a function of $m_Q/m$ for $SU(2)$ and $SU(3)$ with $F=1$. The critical point occurs at $m_Q/m \approx 0.576511$ and $0.398853$, respectively. Right: Value of the critical point $m_{Q,0}/m$ as a function of the number of colors $N$. The points correspond to analytic and numeric determinations of the critical points whereas the blue curve shows the large $N$ prediction in Eq.~\eqref{eq:lNmqm0}.}
\label{fig:minPotSU23}
\end{figure}

For $N=2,3$ the minimum of the potential can be found analytically for $\bar\theta = \pi$.\footnote{Note that for $m_Q < m$ the potential is unbounded in the $\delta_f = 0$ and $x\rightarrow 0$ direction, i.e. far out in moduli space. However, there are higher-order AMSB contributions that among other things generate a squark mass $m^2 |Q_f|^2+ m^2 |\tilde{Q}_f|^2$ that stabilize the potential. These higher-order terms do not significantly affect the local minimum that we determine in this section, which is why we neglect them here. We also checked that the local minimum is much deeper than the potential value in the unbounded direction at the same distance from the origin of moduli space.} We observe the existence of a critical value for the SUSY quark mass $m_{Q, 0}$ below which there exists a unique CP-conserving vacuum and above which two degenerate vacua appear, which break CP spontaneously. For $SU(2)$ and $SU(3)$, the explicit values of the critical point are given by (see App.~\ref{sec:potandmin} for the derivation)
\begin{equation}
	 \left. \frac{m_{Q,0}}{m} \right|_{N=2} = \frac{3}{2} \cos\left( \frac{1}{3} \arccos \left(-\frac{25}{27}\right) -\frac{2\pi}{3} \right) \approx 0.576511 \,
 \label{eq:su2mqmk0}
\end{equation}
and
\begin{equation}
	 \left. \frac{m_{Q,0}}{m} \right|_{N=3} = \frac{14}{9} \cos\left( \frac{1}{3} \arccos \left(-\frac{289}{343}\right) -\frac{2\pi}{3} \right)- \frac{1}{9} \approx 0.398853 \,.
 \label{eq:su3mqmk0}
\end{equation}
In the large $N$ limit, we can also find an analytic expression for $m_{Q,0}/m$ which to leading order is of the form 
\beq
\left. \frac{m_{Q,0}}{m} \right|_{N \gg 1} = \frac{9}{7}\frac{1}{N} + \mathcal O \left( \frac1N \right)^2 \,.
\label{eq:lNmqm0}
\eeq
For $N>3$ the minimum of the potential and the critical value $m_{Q,0}$ can only be found numerically. However, as can be seen in the right panel of Figure~\ref{fig:minPotSU23}, the critical point quickly approaches the large $N$ prediction even for relatively small values of $N$. In fact, the analytic values of the critical point for $SU(2)$ and $SU(3)$ deviate by only 12\% and 7\% from the large $N$ estimate in Eq.~\eqref{eq:lNmqm0}, respectively.

The appearance of the two degenerate minima can be understood by studying $\delta_f$, i.e. the phase of $\phi$, at the minimum. For $m_Q \ll m \sim m_{Q,0}$ and $x\sim m$, the typical location of the minimum, the next to last term in Eq.~\eqref{eq:vlf1} dominates the $\delta_f$ dependence of the potential. Minimizing this term yields $\delta_f = \pi/2 + \pi \ell$ for $\bar\theta = \pi$. For a sufficiently small $m_Q$ this turns out to be the global minimum. Thus the angular degree of freedom of the meson, which can be identified with the $\eta'$, the would-be Goldstone boson of the anomalous $U(1)_A$ symmetry, exactly aligns with the $\bar\theta$ angle which for $\bar\theta = \pi$ leads to a real meson VEV. This behavior can be observed in the left panel of Figure~\ref{fig:minPotSU23} which shows $\delta_f$ at the minimum for $SU(2)$ and $SU(3)$ as a function of $m_Q/m$. Once $m_Q \sim m$ there is a nontrivial interplay between the different terms in the potential and the minima bifurcate at the critical point, leading to two degenerate vacua with complex-valued meson VEVs, implying that CP is spontaneously broken.

\begin{figure}[h]
\centering
  \includegraphics[width=0.47\textwidth]{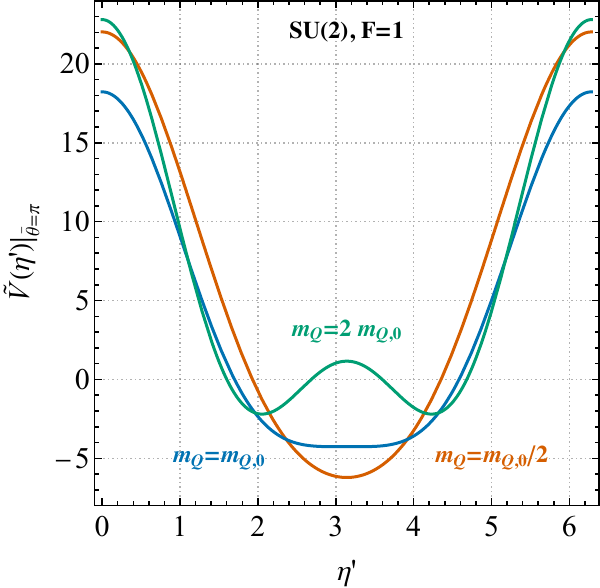}
  \hfill
    \includegraphics[width=0.482\textwidth]{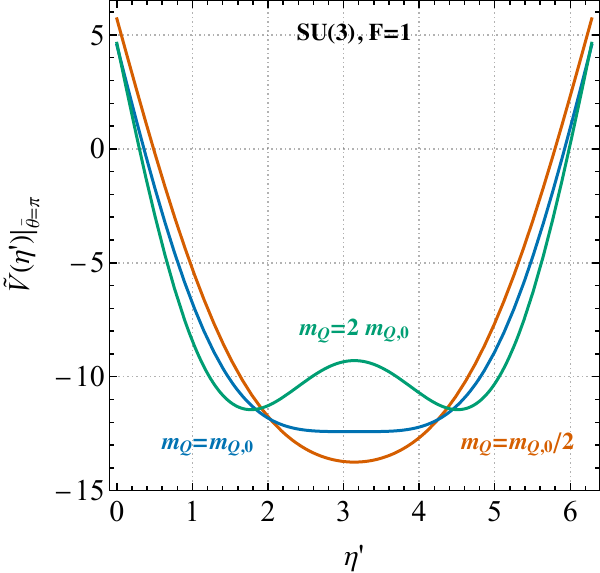}
\caption{Shape of the dimensionless $\eta' = \arg M = 2\delta_f$ potential defined as $\tilde{V}(\eta') = V(\eta')/(\Lambda^{5/2}m^{3/2})$ and $\tilde{V}(\eta') = V(\eta')/(\Lambda^{8/3}m^{4/3})$ for $SU(2)$ (left) and $SU(3)$ (right), respectively. We show the potential for several values of $m_Q$ close to the critical point. At the critical value $m_Q=m_{Q,0}$ the potential is flat around the minimum such that the $\eta'$ mass vanishes, signaling a second-order PT.}
\label{fig:PotSU23}
\end{figure}

The fact that the degenerate CP-breaking minima are continuously connected to the CP-conserving minimum through smooth variations of $m_Q$, as can be seen in the left panel of Figure~\ref{fig:minPotSU23} for $SU(2)$ and $SU(3)$, signals that this is a second-order phase transition. This can also be seen in Figure~\ref{fig:PotSU23} where we show the $\eta'$ potential, i.e. the angular degree of freedom of the meson $\eta' = \arg M = 2\,\delta_F$, where we absorb the decay constant into the definition of $\eta'$ to make it dimensionless. For $m_Q < m_{Q,0}$ there is a single minimum at $\eta'=\pi$ which broadens as the critical point is approached. At $m_Q = m_{Q,0}$ the potential is flat at $\eta'=\pi$ implying that $\eta'$ is exactly massless and for $m_Q > m_{Q,0}$ there are two degenerate minima which are symmetric around $\pi$. Note that the $SU(2)$ potential is smooth at $\eta'=0$ and $\eta'=2\pi$ and trivially $2\pi$-periodic. The $SU(3)$ potential on the other hand is not trivially $2\pi$-periodic. The periodicity is recovered once the second $\ell=1$ branch is included. However, this comes at the cost of a cuspy, non-differentiable point where the branches meet.

The expressions for $f$ (or $x$) and positions of the degenerate vacua $\delta_f$ at finite $N$ are long and uninformative. However, for large $N$ and $m_Q > m_{Q,0}$, they take the simple form
\begin{equation}
\begin{aligned}
&x|_{N \gg 1} = \frac{3m + \Delta}{2} + \mathcal O\left( \frac1N \right) \\
&\delta_f |_{N \gg 1} = \frac\pi2 \pm \left(\frac\pi2 -\frac{(3m + 2m_Q)(3m + \Delta)}{4 m_Q(4m + \Delta)}\frac{\pi}{N} \right) + \mathcal O \left( \frac1N \right)^2\ ,
\label{eq:x_NF1}
\end{aligned}
\end{equation}
with $\Delta \equiv \sqrt{9m^2 - 4mm_Q + 4m_Q^2}$. In the large $N$ limit the vacua approach $\delta_f =0,\pi$ with a VEV $f\gg \Lambda$ for $m,m_Q \ll \Lambda$. In this regime the softly-broken supersymmetric calculation yields reliable results.
Before moving on to a larger number of flavors, let us make a few comments. The phase structure that we find is qualitatively the same as was obtained in~\cite{Gaiotto:2017tne,DiVecchia:2017xpu} for non-supersymmetric one-flavor QCD in the large $N$ limit. However, one should note that our results also apply to small $N$. In fact for $N=2,3$ we even obtain analytic results for the ground state and critical point. The critical point in our discussion has the same $1/N$ scaling as in the non-supersymmetric scenario. However, the scale which determines the location of the critical point is the SUSY breaking scale $m_{Q,0} \sim m\ll \Lambda$ in contrast to $m_{Q,0}\sim \Lambda$ in the non-supersymmetric setup. It would be tempting to extrapolate our results to $m\rightarrow \Lambda$ but in this limit we lose perturbative control over the supersymmetric theory. 
%
\subsection{$N > F \ge 2$: Generalization}\label{subsec:NlargerF2}
%
Despite the absence of full analytic expressions for $F\ge2$, we can still determine numerically  whether CP is spontaneously broken given any quark masses and SUSY breaking scale for $F<N$. 
However, even without resorting to numerical computations, one can establish that at $\bar\theta = \pi$
\begin{itemize}
\item CP is always spontaneously broken when all quark masses are equal
\item  CP is conserved if at least one of the quarks is massless 
\item The effects of heavy quarks decouple from the low-energy as expected in effective theories.
\end{itemize}

While the others are quite obvious, the first claim is non-trivial.
In Section~\ref{subsec:vacuum} we saw that for $F>2$ and equal quark masses both limiting cases $m_Q \gg m$ and $m_Q \ll m$ lead to two degenerate vacua which break CP spontaneously. In order to study the intermediate region $m_Q \sim m$ we either have to minimize the potential numerically or treat the theory in the large $N$ limit. In either case we can make the ansatz that for equal SUSY quark masses $m_Q = m_Q \delta_{f f'}$ the meson VEV is of the form  $M_{f f'} = \phi^2 \delta_{f f'}$ where just as in the $F=1$ case $\phi = f \exp (i\delta_f)$.\footnote{Note that in~\cite{Luzio:2022ccn} it was shown that at the minimum the meson matrix is proportional to the identity $\langle M_{ff'}\rangle \propto \delta_{f f'}$.} The corresponding scalar potential is of the form 
\begin{equation}
\begin{aligned}
V_\ell &= 2 x^{\frac{F-N}{N}} |\Lambda|^{\frac{3N-F}{N}} \bigg(F^2m_Q^2 + F^2 x^2 - F m m_Q \cos(2\delta_f) \\
&\hskip 3em - m(3N-F)x \cos\left[\frac{2F\delta_f - \bar\theta - 2\pi\ell}{N-F} \right] - 2Fm_Q x \cos\left[\frac{2N\delta_f - \bar\theta - 2\pi\ell}{N-F}\right] \bigg),
\end{aligned}
\label{eq:VellFN}
\end{equation}
where $\bar\theta = \theta + F\theta_Q$ and
\beq
x = |\Lambda| \left(\frac{|\Lambda |}{f}\right)^{\frac{2N}{N-F}}\,.
\eeq

For a small number of colors $N$ the minimum has to be found numerically. This is just a special case of the non-degenerate mass scenario for which we will present numeric results for $N=3$ and $N=4$ momentarily. 
However, in the large $N$ limit we can find an analytic expression for the minimum along the same lines as for the $F=1$ case
\begin{equation}
\begin{aligned}
&x|_{N \gg 1} = \frac{3m + \Delta}{2} + \mathcal O\left( \frac1N \right) \\
&\delta_f |_{N \gg 1} = \frac\pi2 \pm \left(\frac\pi2 -\frac{(3m + 2m_Q)(3m + \Delta)}{4 m_Q(4m + \Delta)}\frac{\pi}{N} \right) + \mathcal O \left( \frac1N \right)^2\ ,
\label{eq:x_N}
\end{aligned}
\end{equation}
where again $\Delta \equiv \sqrt{9m^2 - 4mm_Q + 4m_Q^2}$. 
Note that the degenerate vacua persist for all $m_Q>0$, establishing that CP is always spontaneously broken for equal quark masses in large $N$ and $F>1$.

Let us now move beyond the case of degenerate quark masses. In such a scenario the meson matrix is still diagonal but with non-homogeneous values, i.e. $M_{ff'} = \phi_f^2 \delta_{ff'}$, where we again parameterize each $\phi_i$ as $\phi_i = f_i \exp (i\delta_{f_i})$. 
The corresponding potential for $F < N$ is given by
\begin{equation}
\begin{aligned}
V_\ell &= 2\sum_i^F\Bigg[ \frac{1}{f_i^2}\left( \frac{|\Lambda|^{3N-F}}{\prod^F_j f_j^2} \right)^{\frac{2}{N-F}} + f^2_i[m_{Q_i}^2 - mm_{Q_i} \cos(2\delta_{f_i})] \\
&\hskip 3em - \frac1F m(3N - F)\left( \frac{|\Lambda|^{3N-F}}{\prod^F_j f_j^2} \right)^{\frac{1}{N-F}} \cos\left( \frac{2}{N-F}\sum^F_j \delta_{f_j} - \frac{\bar{\theta} + 2\pi\ell}{N-F}\right) \\
&\hskip 5em - 2m_{Q_i} \left( \frac{|\Lambda|^{3N-F}}{\prod^F_j f_j^2} \right)^{\frac{1}{N-F}} \cos\left(2\delta_{f_i} + \frac{2}{N-F}\sum^F_j \delta_{f_j} - \frac{\bar{\theta} + 2\pi\ell}{N-F}\right) \Bigg],
\end{aligned}
\label{eq:Vell32}
\end{equation}
where we used the K\"ahler potential $K = 2 \sum_i \phi_i \phi_i^\dagger$.
In general this potential has to be minimized numerically. Once the minimum is found we can assess if CP is spontaneously broken at $\bar\theta=\pi$.
\begin{figure}[h]
\centering
  \includegraphics[width=0.6\linewidth]{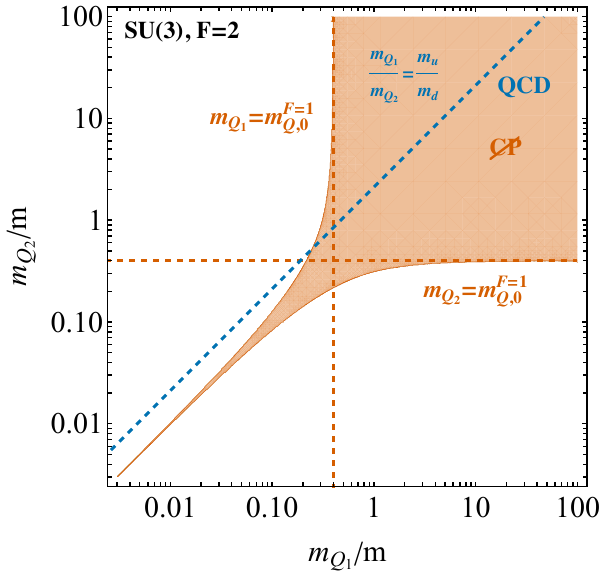}
\caption{Phase diagram at $\bar\theta = \pi$ for $F=2$ and $N=3$ with $|\Lambda|/m = 100$. 
The shaded region shows where CP is spontaneously broken. For $m_{Q_1} / m \gg 1$ and $m_{Q_2}/m \gg 1$ the phase boundary approaches the one-flavor critical point, which is shown as dashed orange lines. The dashed blue line shows where the quark mass ratio agrees with the up and down quark mass ratio in the SM.}
\label{fig:n3f2pd}
\end{figure}

The simplest example with $F>1$ is $SU(3)$ with two flavors $F=2$. This example is especially interesting since the leading-order potential in the small but equal mass limit vanishes as we found in Section~\ref{subsec:SUSYBreakingStabilization}. In order to determine if CP is broken at $\bar\theta =\pi$, a numerical computation or higher-order contributions to the potential are required. A numerical computation reveals that, as expected, also for $F=2$ CP is always spontaneously broken in the equal mass limit. As can be seen in Figure~\ref{fig:n3f2pd} for $|\Lambda|/m = 100$ there is a thinning funnel along the diagonal along which CP is spontaneously broken. This shape can be understood with the help of a perturbative solution to the minimization equations. If we parameterize $m_{Q_{1/2}} = (1 \pm \delta) m_Q$ with $m_Q/m, \delta \ll 1$ the degeneracy of the vacuum w.r.t. $\delta_{f_1}$ and $\delta_{f_2}$ is lifted at order $\mathcal{O}(m_Q^2)$ and $\mathcal{O}(m_Q \delta)$, respectively. To leading order in $\delta$ and $m_Q/m$ it holds that $\delta_{f_1} = \delta_{f_2} \equiv \delta_f$. Schematically we find 
\beq
V/(m^{5/3}\Lambda^{7/3}) \supset \frac{m_Q}{m} \left( -a\, \delta \cos\left( 2\,\delta_f \right) + b\, \frac{m_Q}{m} \cos\left(4\,\delta_f\right)\right)\,,
\eeq
where $a$ and $b$ are positive $\mathcal{O}(1)$ numbers. If $m_Q/m \gg \delta$, i.e. close to the equal mass limit, the potential is minimized at $\delta_f = \pi/4,\, 3\pi/4$, leading to two degenerate vacua with complex meson VEVs that break CP spontaneously. If, however, $\delta \gg m_Q/m$ the potential is minimized at $\delta_f = 0$, i.e. there is a unique vacuum with real meson VEVs and CP remains unbroken. The width $\delta$ of the band in which CP is broken scales with $m_Q/m$ and becomes thinner for smaller masses. 

For larger masses the band where CP is spontaneously broken widens until we reach $m_{Q_1},m_{Q_2} \geq m_{Q,0}^{F=1}$ where CP is always spontaneously broken. This can be understood by considering the limit where one of the masses is much larger than the SUSY breaking scale, e.g. $m_{Q_1}/m \gg 1$. In this limit one can effectively integrate out the heavier flavor such that one obtains the theory with one less flavor, i.e. $SU(3)$ with $F=1$. As we saw in Section~\ref{subsec:F1Results}, the one-flavor theory has a critical mass $m_{Q,0}^{F=1}$ above which CP is spontaneously broken. The boundary in the $m_{Q_2}$ direction approaches asymptotically this critical value for $m_{Q_1}/m \gg 1$ and the same is true for the opposite mass hierarchy.

In Figure~\ref{fig:n3f2pd} we also show in dashed blue the line which corresponds to the ratio of up and down quark masses in the SM. Note, however, that in order to recover real-world QCD we would have to take $m \gtrsim \Lambda$. While the phase diagram depends only weakly on $|\Lambda | / m$, the calculation breaks down when $m\sim |\Lambda |$, implying that any results of such an extrapolation are not reliable.  
\begin{figure}[t]
\centering
  \includegraphics[width=0.48\linewidth]{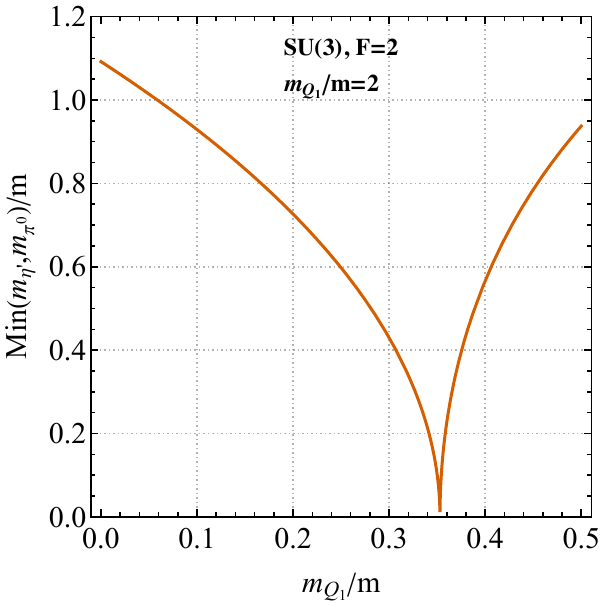}\hfill
    \includegraphics[width=0.48\linewidth]{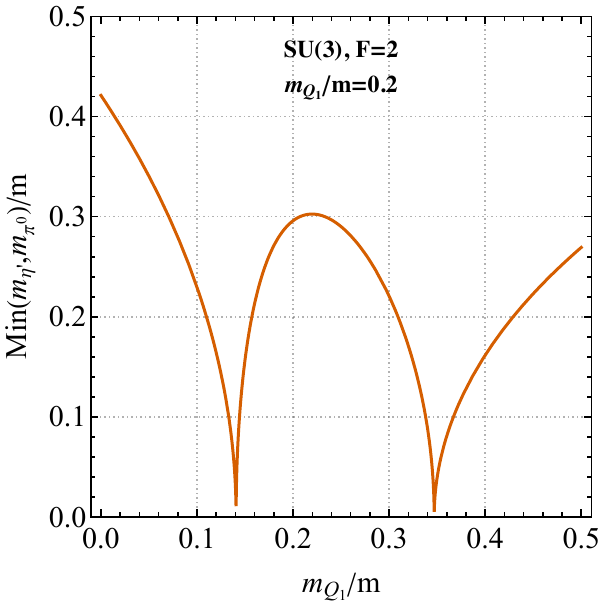}
\caption{The lightest mass eigenvalue of $\eta'-\pi^0$ system for $F=2$, $N=3$, and $|\Lambda|/m = 100$ with $m_{Q_2}/m$ fixed at 2 (left) and 0.2 (right). One linear combination of $\eta'$ and $\pi^0$ becomes massless at the phase boundary, signaling a second-order phase transition. }
\label{fig:eigmn3f2}
\end{figure}
We can also explicitly verify that the boundary between the CP conserving and breaking regions corresponds to a second-order phase transition. In Figure~\ref{fig:eigmn3f2}, we plot the lightest mass eigenvalue of two neutral GBs $\eta'$ and $\pi^0$, where $\eta'$ is the would-be GB of the anomalous $U(1)_A$ symmetry and $\pi^0$ the pNGB along the Cartan generator direction of the spontaneously broken $SU(2)_A$ symmetry. In both cases we take $|\Lambda|/m = 100$ and fix one of the quark masses to $m_{Q_2}/m = 2$ (left panel) and $m_{Q_2}/m = 0.2$ (right panel). For $m_{Q_2}/m = 2$~$(0.2)$ we cross the phase boundary once (twice) and as can be seen there is a massless degree of freedom at the crossing point, signaling a second-order phase transition.
\begin{figure}[t]
\centering
  \includegraphics[width=0.48\linewidth]{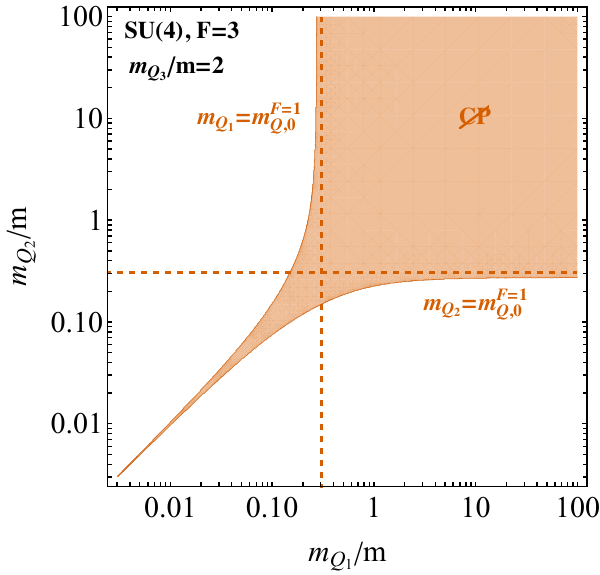}\hfill
        \includegraphics[width=0.48\linewidth]{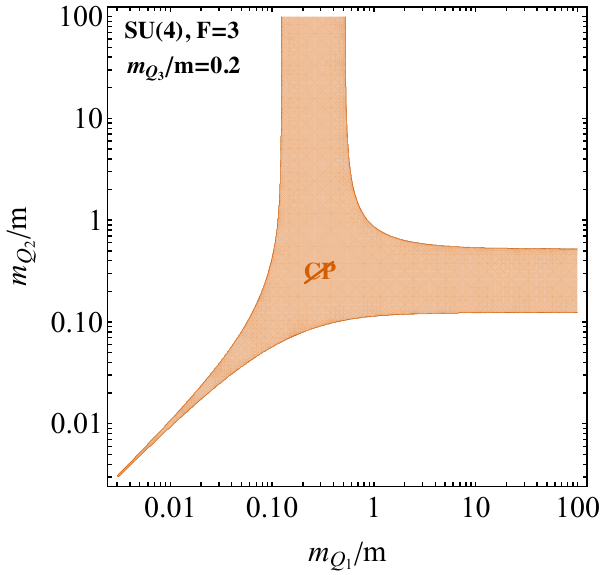}
\caption{Phase diagram at $\bar{\theta} = \pi$ for $F=3, N =4$ and $|\Lambda|/m = 100$ with $m_{Q_3}/m$ fixed at 2 (left) and 0.2 (right). The shaded area shows where CP is spontaneously broken and the dashed lines in the left panel correspond to the one-flavor critical mass value.}
\label{fig:n4f3}
\end{figure}

The generalization to $F\geq 3$ is straightforward. There will always be a region around the degenerate mass line in which CP is spontaneously broken. This region will widen up until all masses are above the SUSY breaking scale in which case CP is always spontaneously broken. As a concrete example, let us consider $F=3$ and $N=4$. In Figure~\ref{fig:n4f3} we show slices of the three-dimensional parameter space where one of the quark masses $(m_{Q_3})$ is held fixed at $m_{Q_3}/m = 2$ at the left panel and $m_{Q_3}/m =0.2$ in the right panel. If the third quark mass is sufficiently above the SUSY breaking scale it does not play a significant role in determining the phase structure anymore. This can be seen in the left panel which qualitatively reduces to the $F=2$ case in Figure~\ref{fig:n3f2pd}. However, the asymptotic values of the phase boundaries do not agree with the critical value $m_{Q,0}^{F=1}$ since $m_{Q_3} \sim m$ is not much larger than the SUSY breaking scale.

If, however, the third quark mass is light CP is restored at large values along the $m_{Q_1} = m_{Q_2}$ line. This can be understood by realizing that when $m_{Q_1}/m = m_{Q_2}/m \gg 1$ the two heavy flavors can be integrated out and one obtains the theory with $F=1$ which has a critical point at $m_{Q,0}/m \approx 0.4$. If the remaining light quark mass is below the critical value CP is restored in this limit as can be seen in the right panel of Figure~\ref{fig:n4f3}.

\subsection{$F \ge N$: To Infinity and Beyond}
\label{sec:FgtrNCP}
Once $F$ is increased beyond $N-1$ the ADS superpotential no longer describes the low-energy dynamics of the theory~\cite{Taylor:1982bp,Seiberg:1994bz,Seiberg:1994pq}. For $F=N$ the anomaly matching conditions are satisfied by a confining theory with color neutral meson and baryon fields $M_{f f'}$ and $B, \bar{B}$, respectively.\footnote{The baryon fields are completely antisymmetric color singlet combinations of the quark superfields $Q$ and $\bar{Q}$, i.e. $B = \epsilon^{f_1 \cdots f_F} \epsilon_{a_1 \cdots a_N} Q_{f_1}^{a_1} \cdots Q_{f_F}^{a_N}$, where $a_i$ are color indices.} For $F=N$ there is also a quantum modified constraint on the moduli space of the form $\det(M) - \bar{B} B = \Lambda^{2N}$. We first focus on $N>2$ and will comment on $N=2$ at the end. The quantum modified constraint is implemented in the superpotential in terms of a Lagrange multiplier field $X$
\beq
W = X\left( \frac{\det (M) - \bar{B} B}{\Lambda^{2N}} - 1 \right) +  \text{Tr}(m_Q M)\,.
\label{eq:fnsp}
\eeq
Note that at the meson point where $B = \bar{B} = 0$, already the quantum modified constraint breaks $U(1)_A$. In addition the mesons are uncharged under $U(1)_R$, such that in contrast to the previous sections the $\eta'$ mass is not protected in this scenario and the $\eta'$ obtains a mass of order $\Lambda$. This scenario therefore resembles non-supersymmetric QCD.

In the $m\rightarrow 0$ limit there are $N$ degenerate vacua
\beq
\begin{split}
\langle B\rangle =\langle \bar{B}\rangle =0\,,\qquad \langle X\rangle =-\left(\det (m_Q)\right)^{1/N} |\Lambda|^2 e^{i\frac{\bar{\theta}+2\pi k}{N}}\,,\\
\quad \left\langle M_{f f'} \right\rangle = - \langle X\rangle \, (m_Q^{-1})_{ff'} \,, \qquad k=0,\ldots, N-1\,.
\end{split}
\eeq
It is important to stress that in this limit the VEV of the meson matrix $M$ is exact since its form is fixed by holomorphy and symmetry arguments. Adding a small amount of SUSY breaking the degeneracy of the vacua is lifted
\beq
\delta V_{\rm AMSB} = -2(2N + 1) (\det (m_Q))^{1/N} m\, |\Lambda |^2 \cos\left(\frac{\bar{\theta} + 2\pi k}{N}\right)\,.
\eeq
For general $\bar\theta$ this has a unique minimum. However, for $\bar\theta = \pi$ there are two degenerate minima with $k=0,N-1$, such that the meson VEV is complex and CP is always spontaneously broken.

Note that the opposite limit $m\gg m_Q$ cannot be taken reliably. The theory is inherently strongly coupled and due to SUSY breaking the meson VEV is not required to be a holomorphic function of the potential parameters. In fact the existence and location of global and local minima depends on incalculable prefactors in the K\"ahler potential as well as higher-order terms in the K\"ahler potential which are not suppressed when the VEVs are of order $\Lambda$. As was shown in~\cite{Murayama:2021xfj,Csaki:2022cyg} it cannot even be determined if the vacuum is at the meson point, where $M\neq 0$ and $B=\bar{B}=0$ or the baryon point where $M=0$ and $B,\bar{B}\neq 0$. For this reason we cannot ascertain whether the vacuum at $\bar{\theta}=\pi$ spontaneously breaks CP.

Let us now comment on the special case of $F=N=2$. The fundamental representation of $SU(2)$ is real, such that quarks and anti-quarks transform in the same representation. Hence the flavor symmetry is enhanced to $SU(4)$ with the holomorphic gauge-invariant and anti-symmetric meson matrix being given by
\beq
M_{f f'} = \epsilon_{a_1 a_2} Q^{a_1}_{f} Q^{a_2}_{f'}\,,\quad f,f'=1,\ldots,4\,,
\eeq
for which the quantum modified constraint takes the form Pf$(M) = \Lambda^4$, where Pf$(M)$ stands for the Pfaffian of $M$. This constraint is again implemented in the superpotential using a Lagrange multiplier field $X$
\beq
W = X \left( \frac{\text{Pf}(M)}{\Lambda^4} -1 \right) + \frac{1}{2}\text{Tr}(m_Q M)\,,
\eeq
where also $m_Q$ is a $4\times 4$ matrix in this case. The supersymmetric minimum for $m\rightarrow 0$ is again exact and is given by
\beq
\langle X\rangle  = -\left(\text{Pf}(m_Q)\right)^{1/2} |\Lambda|^2 e^{i\frac{\bar{\theta}+2\pi k}{2}}\,,\quad \langle M_{ff'} \rangle = -\langle X \rangle (m_Q^{-1})_{ff'}\,,\quad k=0,1\,.
\eeq
Thus there are two degenerate minima. Introducing a small amount of SUSY breaking $m\ll m_Q$ we get a correction to the potential of the form
\beq
\delta V_{\rm AMSB} = -12 m \left(\text{Pf}(m_Q)\right)^{1/2} |\Lambda|^2 \cos\left( \frac{\bar{\theta} + 2\pi k}{2}\right)\,.
\eeq
For generic $\bar{\theta}$ this lifts the degeneracy. However, for $\bar{\theta}=\pi$ $\delta V_{\rm AMSB}$ vanishes and the degeneracy is not lifted. Thus for small SUSY breaking CP is spontaneously broken for $N=2$ as well. In the reverse limit, just as for the $N>2$ case, a reliable computation is not possible.

For $F=N+1$ and $N>2$ the baryons are in the fundamental representation of the flavor group $SU(F)$ and the classical and quantum constraints on the moduli space are identical and follow from the superpotential
\begin{equation}
W = \frac{B M \bar{B} - \det (M)}{\Lambda^{2N-1}} + \text{Tr}(m_Q M)\,.
\end{equation}
As a first step we can again consider the two limiting cases $m_Q \gg m$ and $m\gg m_Q$. In the supersymmetric limit $m\rightarrow 0$ the vacuum can be determined exactly and has the form
\beq
\langle B\rangle =\langle \bar{B}\rangle = 0\,,\quad \left\langle M_{ff'}\right\rangle = (m_Q^{-1})_{ff'} \left(|\Lambda|^{2N-1} \det (m_Q)\right)^{1/N} e^{i\frac{\bar{\theta +2\pi k}}{N}}\,,\quad k=1,\ldots ,N-1\,, 
\eeq
where we again obtain $N$ degenerate vacua. The degeneracy is lifted by SUSY breaking effects which give a contribution to the potential of the form
\beq
\delta V_{\rm AMSB} = -6N\, m \left( |\Lambda |^{2N-1} \det (m_Q)\right)^{1/N}  \cos\left(\frac{\bar{\theta} + 2\pi k}{N}\right)\,,
\eeq
which has two degenerate minima for $\bar\theta = \pi$ and breaks CP spontaneously. In the opposite limit $m_Q \ll m \ll \Lambda$, the minimum is determined by the SUSY breaking contributions to the potential. Just like in the $F<N$ case the theory has an exact $SU(F)_L \times SU(F)_R$ flavor symmetry in the $m_Q\rightarrow 0$ limit. Thus we make the ansatz $M = f^2 e^{2i \delta_f} U$, where $U$ is an arbitrary $SU(F)$ matrix which is a flat direction in the massless limit. With this ansatz and assuming a canonical K\"ahler with arbitrary prefactor $K = \text{Tr} (M^\dagger M)/(\alpha |\Lambda|^2)$ the scalar potential takes the form
\beq
V = \alpha (N+1) \frac{f^{4 N}}{|\Lambda|^{4N-4}} -2\frac{m}{|\Lambda |^{2N-1}} (N-2) f^{2N+2} \cos\left( (2N+2)\delta_f -\bar{\theta}\right)\,.
\eeq
This is solved by
\beq
f^2 = \left(\frac{m}{|\Lambda|}\frac{N-2}{\alpha N}\right)^{1/(N-1)} |\Lambda|^2\,,\quad 2\delta_f = \frac{\bar{\theta}}{N+1}\,.
\eeq
Note that there is only one inequivalent solution for $\delta_f$ as a shift of $2\delta_f$ by $2\pi k/(N+1)$ can be absorbed into $U$, since $e^{2\pi 
 i k/(N+1)}$ is in the center of $SU(F)$. Also note that for $\alpha \sim \mathcal{O}(1)$ we find that $f^2 \ll |\Lambda |^2$ such that higher-order terms in the K\"ahler potential are suppressed and the calculation is reliable as long as $m\ll |\Lambda |$. 
 
Just like in the $F<N$ case that we discussed in Section~\ref{subsec:SUSYBreakingStabilization}, the introduction of quark masses breaks the $SU(F)_L\times SU(F)_R$ flavor symmetry explicitly and therefore lifts the flat direction and fixes the $SU(F)$ matrix $U$ in the parameterization of the meson VEV. We can always diagonalize $U$, such that it is of the form as in Eq.~\eqref{eq:Uparam}. Thus to leading order in $m_Q$ the scalar potential gets corrected by
\beq
\delta V_{m_Q} = - 2 \left(\frac{m}{|\Lambda |}\frac{N-2}{\alpha N}\right)^{1/(N-1)} m |\Lambda|^3 \sum_{i=1}^{N+1} m_{Q,i} \cos\left(\frac{\bar{\theta}}{N+1}+\alpha_i\right)\,,
\eeq
which is exactly of the same form as Eq.~\eqref{eq:smallMassPot} for $F<N$, such that the same conclusions hold. In the degenerate mass case there is an unbroken $SU(F)_V$ global symmetry. At $\bar{\theta}=\pi$ this vectorial flavor symmetry is also respected by the meson VEV as $\delta V_{m_Q}$ is minimized in this limit by
\beq
\alpha_1 = \ldots =\alpha_{N+1} = \frac{2\pi k}{N+1}\,,\quad k=0,N\,,
\eeq
which leaves the flavor symmetry unbroken but spontaneously breaks CP. Away from the degenerate limit it becomes more favorable to cancel the argument of the cosines corresponding to the largest quark masses. Due to the restriction that $U$ is an $SU(F)$ element we cannot cancel all arguments. The resulting phase structure depends on the mass hierarchies and cannot be found in full generality and typically one has to resort to numeric methods. 
\begin{figure}[t]
\centering
  \includegraphics[width=0.45\linewidth]{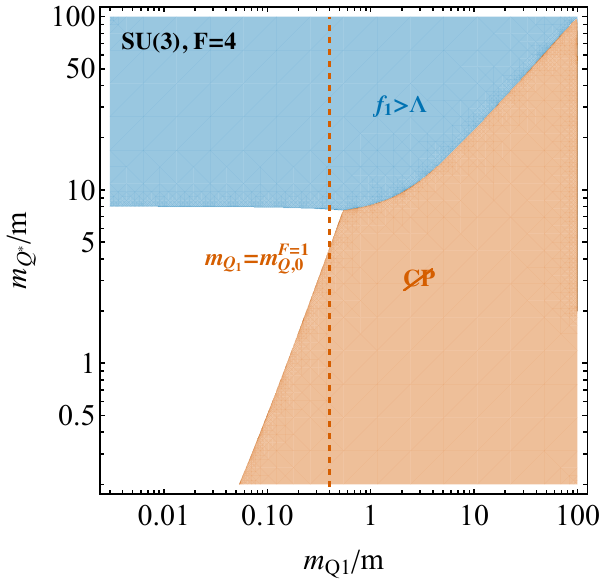}\hfill
    \includegraphics[width=0.45\linewidth]{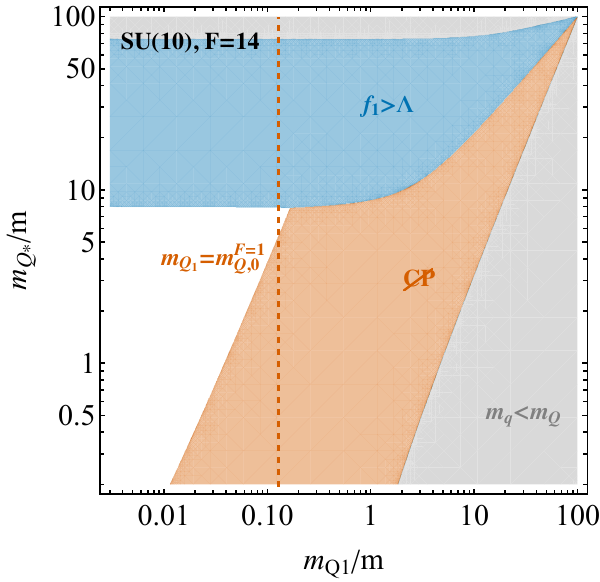}
\caption{Phase diagram at $\bar{\theta} = \pi$ for $F=4, N=3$ (left) and $F=14, N=10$ (right). $\alpha=1$,  $|\Lambda|/m = 100$ and all but one quark have equal masses $m_{Q^*}$. The orange area shows where CP is spontaneously broken and blue indicates that the theory is no longer trusted because $f_1 > \Lambda$. The region shaded in gray in the right plot, corresponds to parameter values where the dual quarks are lighter than the mesons and cannot be integrated out. In this region the effective superpotential in Eq.~\eqref{eq:NgtrFp1Weff} does not provide an accurate description of the theory. The dashed lines correspond to the one-flavor critical mass value.} 
\label{fig:n34f45}
\end{figure}
In the left panel of Figure~\ref{fig:n34f45} we show numeric results for one specific case with $SU(3)$ and $F=4$. We take $|\Lambda| /m =100$ and fix $m_{Q_2}=m_{Q_3} = m_{Q_4}\equiv m_{Q^*}$. The orange shaded region shows where CP is spontaneously broken, whereas the blue region denotes where one of the meson VEVs exceeds $|\Lambda|$, indicating that higher order terms in the K\"ahler potential become relevant, implying that we cannot trust the calculation anymore. When $m_{Q^*} \gg m_{Q_1}$ we can integrate out the $N$ heavy quarks and are left with $SU(3)$ and $F=1$. The resulting one-flavor theory has a critical point for $m_{Q_1}$ below which CP is conserved and above which it is spontaneously broken. The dashed line shows the critical point in the $F=1$ theory which should be reached when $m_{Q*}\rightarrow \Lambda$ fully decouples. Also note that the orange shaded region extends all the way to (but excluding) zero in the vertical $m_{Q^*}$ direction. This can be understood by noticing that when $m_{Q_1} \gg m_{Q^*}$ the first quark decouples leaving an effective theory with $N$ mass-degenerate light quarks, for which CP is always spontaneously broken. 

Before moving on, let us consider the special case of $N=2$ and $F=3$. Quarks and anti-quarks transform in the same representation and the flavor symmetry is enhanced to $SU(6)$. The meson matrix is just like in the $F=N=2$ case anti-symmetric $M_{ff'} = - M_{f'f}$ with $f=1,\ldots ,6$. The classical constraint on the moduli space is encoded in the superpotential
\beq
W = -\frac{1}{\Lambda^3} \text{Pf}(M) + \frac{1}{2}\text{Tr} (m_Q M)\,.
\eeq
The supersymmetric limit $m\rightarrow 0$ is straightforward and one finds two degenerate minima
\beq\label{eq:minN2F3}
M_{f f'} = \left(\text{Pf}(m_Q)\right)^{1/2} |\Lambda|^{3/2} e^{i\frac{\bar{\theta}+2\pi k}{2}} (m_Q^{-1})_{f f'}\,,\quad k=0,1\,.
\eeq
Adding a small amount of SUSY breaking lifts the degeneracy for generic $\bar{\theta}$
\beq
\delta V_{\rm AMBS} = -12 m \left(\text{Pf}(m_Q)\right)^{1/2} |\Lambda |^{3/2} \cos\left(\frac{\bar{\theta} + 2\pi k}{2}\right)\,.
\eeq
However, for $\bar{\theta} = \pi$ the degeneracy is not lifted. In fact $\delta V_{\rm AMSB}$ vanishes. Thus also for $N=2$ CP is spontaneously broken in the $m_Q \gg m$ limit. The opposite limit is much more delicate. The theory is classically scale invariant in the $m_Q\rightarrow 0$ limit, i.e. there is no tree-level AMSB potential. The positive 2-loop mass contribution to the scalar potential drives the meson VEV to the origin of moduli space, where the theory confines without chiral symmetry breaking (see also~\cite{deLima:2023ebw}). Adding a mass perturbation will at some point misalign the minimum from the origin since it has to be continuously connected to the large $m_Q$ limit. As the AMSB contribution to the potential only stabilizes the minimum at the origin, the minimum will be misaligned by $\mathcal{O}(m_Q)$ contributions, leading to a linearized form of Eq.~\eqref{eq:minN2F3}. And just as before the $\mathcal{O}(m_Q m)$ contributions will not break the degeneracy of the vacuum, such that the conclusions from the $m_Q\gg m$ limit should still hold.

For $N+1 < F < 3/2 N$ the theory can be studied in a weakly-coupled dual description with gauge group $SU(F-N)$ that has a dynamical scale $\tilde{\Lambda}$. The low-energy spectrum contains the meson matrix $M$ and $F$ quarks $q_i$ and anti-quarks $\bar{q}_i$ in the fundamental and anti-fundamental representation of $SU(F-N)$, respectively. The dynamics is encoded in the following superpotential
\begin{equation}
W_d = \frac{1}{\mu} q_i M_{ij} \bar{q}_j +  \text{Tr}(m_Q M)\,,
\end{equation}
where $\mu$ is a scale which is needed to convert between the dynamic scale of the electric and magnetic theory
\beq
\Lambda^{3N-F} \tilde{\Lambda}^{3\tilde{N}-F} = (-1)^{F-N} \mu^F\,,
\eeq
where $\tilde{N}=F-N$. If $M$ gets a VEV whose magnitude is larger than $m_Q$ the dual quarks get a mass and can be integrated out. The low-energy theory is a pure $SU(F-N)$ gauge theory with dynamical scale $\tilde{\Lambda}_{\rm eff}$ which is related to the original scale through the scale matching relation
\beq
\left(\frac{\tilde{\Lambda}_{\rm eff}}{\tilde{m}}\right)^{3\tilde{N}} = \left( \frac{\tilde{\Lambda}}{\tilde{m}}\right)^{3\tilde{N}-F}\,,
\eeq
where $\tilde{m}^F = \det(M/\mu)$ is the effective mass of the dual quarks. The low-energy superpotential is therefore generated by gaugino condensation in the pure $SU(N-F)$ gauge group amended by the meson mass term
\begin{equation}\label{eq:NgtrFp1Weff}
    W_d^{\rm eff} = \tilde{N} \tilde{\Lambda}_{\rm eff}^3 + \text{Tr}(m_Q M) = (N-F) \left(\frac{\Lambda^{3N-F}}{\det M}\right)^{\frac{1}{N-F}} + \text{Tr}(m_Q M)\,,
\end{equation}
where we used the scale matching relation to write it as a superpotential in terms of the mesons only. This is exactly the same form as the ADS superpotential for $F<N$ with $N$ and $F$ exchanged. The phase structure is therefore also anologous with the only difference that we do not have full control over the K\"ahler potential in this case, which we take to be canonical for the mesons. However, this implies that results are only reliable if the meson VEV is smaller than $\Lambda$ since otherwise higher-order terms in the K\"ahler potential would become important. We refer the reader to Section~\ref{subsec:NlargerF2} for a discussion of the phase structure as it is completely analogous to the $F<N$ case. Here we only show numeric results for one special case with $SU(10)$ and $F=14$ in the right panel of Figure~\ref{fig:n34f45}, which assumes $F-1$ degenerate quarks with mass $m_{Q^*}$ and one additional quark with mass $m_{Q_1}$. The region where CP is spontaneously broken is shaded in orange and the region where the meson VEV exceeds $\Lambda$ and the perturbative calculation is not reliable is masked in blue. Additionally the region where the dual quark masses are lighter than the meson masses is shaded in grey. For this estimate we assumed that $\mu\sim \Lambda$ which up to an $\mathcal{O}(1)$ number should be a good estimate (see e.g. the discussion in~\cite{Csaki:2011xn}). In this region it is not justified to integrate out only the dual quarks and the derivation of the effective superpotential in Eq.~\eqref{eq:NgtrFp1Weff} is not consistent. The phase structure shows the same qualitative features as all other cases that we have considered. When $m_{Q^*} \gg m_{Q_1}$ the theory reduces to the one-flavor case with a critical point in the $m_{Q_1}$ direction and in the opposite limit the low-energy theory contains $F-1$ degenerate quarks for which CP is always spontaneously broken.

%
\section{Domain wall}\label{sec:DomainWall}
%
The spontaneously broken CP symmetry of the previous section comes with CP domain wall solutions (scalar field configurations which interpolate between the two CP violating vacua). CP domain walls in softly broken SQCD for $F<N$ with degenerate quark masses were thoroughly studied by Patrick Draper in~\cite{Draper:2018mpj}. He showed that there are qualitatively different domain wall solutions depending on the number of colors $N$ and flavors $F$. These solutions differ by how much of the flavor symmetry they break along their trajectory in field space. Classically the theory has a $U(F)_L \times U(F)_R$ flavor symmetry which is spontaneously and explicitly broken to $U(F)_V$ by the meson VEV and quark masses (if they are equal).\footnote{When quantum effects are included, $U(1)_A$ is anomalous and explicitly broken. However, in SQCD for all $F$ but $F=N$ the $U(1)_R$ symmetry mixes with $U(1)_A$, such that there is a non-anomalous axial $U(1)$ symmetry which is explicitly broken only when SUSY is broken. For this reason the $\eta'$ mass in SQCD is of the order of the SUSY breaking scale and not the strong coupling scale.} The corresponding pNGBs are the $\eta'$, the GB along the $U(1)_A$ direction, and the pions $\pi^j$, the GBs along the $SU(F)_A$ direction, and can be parameterized as
\beq
M = |f|^2 e^{i \eta'} e^{i \pi^j T^j}\,,
\eeq
where $T^j$ are the $SU(F)_A$ generators. We also absorbed the $\eta'$ and pion decay constants $f= f_\pi = f_{\eta'}$ into the definition of the fields to make them dimensionless.

Domain walls whose profile is only in the $\eta'$ direction do not break the residual $SU(F)_V$ symmetry, whereas domain walls in the pion directions on the other hand break the $SU(F)_V$ flavor symmetry. It was pointed out in~\cite{Draper:2018mpj} that 
pure $\eta'$ domain walls exist only when $\gcd(N,F) = 1$ and pion-like domain walls are required when $\gcd(N,F) > 1$. 
This can be understood from the explicit form of the AMSB QCD potential. We assume that all quark masses are equal and only include the neutral pions along the $F-1$ Cartan generators $t^j$ of $SU(F)_A$ which are of the form 
$\sqrt{2j(j+1)}\, t^j=\text{diag} (1,\ldots, 1,-j,0,\ldots ,0)$ where the first $j$ diagonal elements are 1 and $j=1, \ldots , F-1$. 
The resulting scalar potential for $\bar{\theta} =\pi$ for the neutral GBs has $N-F$ branches labeled by $l=0,\ldots , N-F-1$ and is given by
\begin{equation}
\begin{aligned}
V_\ell =& -2 x^{\frac{F-N}{N}} |\Lambda|^{\frac{3N-F}{N}} 
\Bigg[ 
 m(3N-F)x \cos\left(\frac{F}{N-F}\eta' - \frac{2\ell+1}{N-F}\pi \right)  \\
& + m \sum_{i=1}^F m_Q \cos \left(\eta' + \sum^{F-1}_{j=1} t^j_i \pi^j \right) + 2 x \sum_{i=1}^F m_Q \cos\left( \frac{N}{N-F}\eta' + \sum^{F-1}_{j=1} t^j_i \pi^j - \frac{2\ell+1}{N-F}\pi \right) \Bigg],
\end{aligned}
\label{eq:VellmQ}
\end{equation}
where
\beq
x \equiv \frac{1}{f^2} \left( \frac{|\Lambda|^{3N-F}}{f^{2F}} \right)^{\frac{1}{N-F}}
\eeq
and $t^j_i$ is the $i$-th diagonal element of the $j$-th Cartan generator.
In the region of masses where CP is spontaneously broken there are two degenerate minima for this potential. One of them usually lies on the $l=0$ branch, while the other could be on any of the other branches, depending on the values of the parameters. In the limit where SUSY is only weakly broken, i.e. $m\ll m_Q$, we saw in Section~\ref{sec:PotMassStabilization} that the second minimum lies on the $\ell = N-F-1$ branch. However, when $m_Q \ll m$ the potential is stabilized by SUSY breaking and the position of the vacua does not have to agree with the supersymmetric theory.

In order to understand the structure of the domain wall in field space it is useful to identify the  discrete gauge symmetries (DGS) of the field configurations, i.e. discrete shift symmetries which leave the potential \Eq{eq:VellmQ} unchanged and lead to physically equivalent field configurations:   
\begin{align}
&\ell \rightarrow \ell + N-F\,,\\
&\pi^j \rightarrow \pi^j + 2\pi\,,\\
&\eta' \rightarrow \eta' + 2\pi, \quad \ell \rightarrow \ell - F\,,\label{eq:dgsetap} \\
&\sum^{F-1}_{j=1} t^j_i \pi^j \rightarrow \sum^{F-1}_{j=1} t^j_i \pi^j + 2\pi/F, \quad \eta' \rightarrow \eta' - 2\pi/F, \quad \ell \rightarrow \ell - 1\,. \label{eq:dgspe}
\end{align}
From these symmetries we can extract that there are $N-F$ branches, where the first and $(N-F)$-th branch are identified. The pion fields have a compact field range with period $2\pi$, whereas a $2\pi$ shift in the $\eta'$ direction moves us from branch $\ell$ to branch $\ell +F$. Finally a combined shift of the pion fields and the $\eta'$ moves us to the adjacent branch. In our basis for the Cartan generators, this shift is performed along the $\pi^{F-1}$ direction whose associated generator $t^{F-1}$ generates the center of $SU(F)$. Thus, a shift along $\pi^{F-1}$ in combination with the $2\pi$-periodicity of the cosine can shift the argument of each cosine in the second line of Eq.~\eqref{eq:VellmQ} by $2\pi /F$.

Trivial domain walls are those where only a single field ($\eta'$ in our case) has a non-trivial profile that does not break any additional flavor symmetries. 
The DGS's in Eq.~(\ref{eq:dgsetap}) are very useful to find a simple condition for when such trivial domain walls may exist. In particular Eq.~(\ref{eq:dgsetap}) tells us that a $2\pi$ shift in $\eta'$ is equivalent to shifting the branches by $F$, thus as we cycle around several times in $\eta'$ space we will keep changing the branches by $F$. The existence of the trivial domain wall then depends on whether we are cycling through all the branches this way. Since there are $N-F$ branches, and we keep jumping by $F$, this will come down to the question whether the greatest common divisor of these numbers is trivial, ie. $\gcd(N-F,F) = \gcd(N,F) = 1$. Hence the condition for ensuring existence of a trivial domain wall made entirely of $\eta'$ is  $\gcd(N,F) = 1$, while for $\gcd(N,F) > 1$ it is possible that such trivial domain walls may not exist. On the other hand, it is clear from Eq.~(\ref{eq:dgspe}) that a simultaneous trajectory  of the $\eta'$ and $\pi^{F-1}$ ($\eta'-\pi^{F-1}$) directions can  traverse all branches, so the associated domain wall is always stable. 

As an example consider $F=2, N=4$, which has two branches ($\ell=0$ and $\ell=1$), and the two degenerate minima are located on different branches. However, one cannot move to a different branch along the $\eta'$ direction because $\ell$ only changes by $F=2$, so the $\eta'$ domain wall is not possible. However for $F=2, N=5$, $\eta'$ can visit any branches since $\gcd(N,F)=1$.

Similar arguments apply to $F>N$ configurations because they have $F-N$ branches, and hence no stable $\eta'$ domain walls are possible when $\gcd(F,N) > 1$. 
In particular, for $N < F < 3N/2$, the relevant GB potential for $\bar{\theta}=\pi$ is given by
\begin{equation}
\begin{aligned}
V_\ell = 
& - 2 (3N - 2F) x^{\frac{F}{N}} |\Lambda|^{\frac{3N-F}{N}} m \cos\left(\frac{F}{F-N}\eta' - \frac{2\ell + 1}{F-N}\pi\right) \\
& - 4 F x^{\frac{F-N}{N}} |\Lambda|^{\frac{3N-F}{N}}  m \sum_{i=1}^{F} m_Q \cos\left(\eta' + \sum^{F-1}_{j=1} t^j_i \pi^j\right) \\
& - 2 \alpha x |\Lambda|^2 \sum_{i=1}^{F} m_Q \cos\left(\frac{N}{F-N}\eta' +\sum^{F-1}_{j=1} t^j_i \pi^j -  \frac{2\ell + 1}{F-N}\pi\right),
\end{aligned}
\label{eq:VFgN}
\end{equation}
which retains DGS's similar to those found for the $F<N$ potential, suggesting the same stability conditions for the existence of domain walls.

For $F=N$, the scalar potential at $\bar\theta = \pi$ around the meson point with $B=\bar{B}=0$ has the form 
\begin{equation}
\begin{aligned}
V = & \alpha N m_Q^2 \Lambda^2 + \beta |\Lambda|^4 + \frac{1}{f^4} \left( \frac{f}{|\Lambda|} \right)^{4N}(\alpha N |X|^2 + \beta f^4 |\Lambda|^2) + 4m|X|\cos(\delta X)\\
& + 2 \beta \left( \frac{f}{|\Lambda|} \right)^{2N} |\Lambda|^4 \cos(N\eta' -\pi) 
-4m f^2 \sum_{i=1}^{N} m_Q \cos\left(\eta' +\sum^{N-1}_{j=1} t^j_i \pi^j \right)\\
& - 2 \alpha \frac{|\Lambda|^2}{f^2} \left( \frac{f}{|\Lambda|} \right)^{2N}\sum^N_i m_Q |X| \cos\left((N-1)\eta' + \sum^{N-1}_{j=1} t^j_i \pi^j - \pi + \delta X\right), \,
\end{aligned}
\label{eq:VFeqN}
\end{equation}
where $X$ is the Lagrange multiplier field of \Eq{eq:fnsp} with $\delta X \equiv \arg X$.
Note that we added a K\"ahler term of the form $X^\dagger X/(\beta |\Lambda|^4)$ for the Lagrange multiplier field. Due to this modification the quantum modified constraint is only satisfied to leading order in the SUSY breaking and also allows non-vanishing quantum fluctuations of the $\eta'$ field, i.e. the phase of the meson matrix. The limit $\beta\rightarrow \infty$ corresponds to making $X$ a non-dynamical Lagrange multiplier. This limit enforces the quantum modified constraint exactly and also makes $\eta'$ infinitely heavy and thus confined to its minimum. For $\beta\sim \mathcal{O}(1)$ the $\eta'$ potential scales as $|\Lambda|^4$, leading to $m_{\eta'} \sim \Lambda$. The reason why $\eta'$ can be heavy in this case is that the mesons are only charged under $U(1)_A$ but not under $U(1)_R$, implying that there is no non-anomalous chiral $U(1)$ that protects the $\eta'$ potential.

There are also no branches, suggesting that the minima can in principle be connected via the $\eta'$ and pion directions. In the limit of small SUSY breaking, i.e. $m \ll m_Q$, we saw in Section~\ref{sec:FgtrNCP} that the two minima are related by a relative phase of $e^{2\pi i/N}$ in the meson matrix $M=e^{i\eta'} U$. This phase can be obtained by a movement in the $\eta'$-direction or in the $\pi^{F-1}$-direction, which connects the identity to $e^{2\pi i/N} \mathbb{1}_N$, an element in the center of $SU(N)$. As the $\eta'$ is heavy for $F=N$ it is beneficial to connect the minima along the pionic direction. This yields a domain-wall tension that scales like $T\sim m_Q |\Lambda |^2$ which is the same scaling as the BPS domain-wall solution in unbroken SQCD, a domain wall solution that preserves half of the supercharges~\cite{Dvali:1996xe,Abraham:1990nz,Cecotti:1992rm}, for unbroken $F=N$ SQCD~\cite{Ritz:2002fm,Ritz:2004mp,Benvenuti:2021com}.\footnote{Note that~\cite{Ritz:2002fm,Ritz:2004mp,Benvenuti:2021com} work in coordinates that enforce the quantum modified constraint exactly, i.e. stable domain walls are only possible in the pionic direction while the $\eta'$ is non-dynamical.}

Incidentally, for $F=N$, there are also baryonic degrees of freedom $B$ and $\bar B$, and domain wall solutions along the baryonic directions exist in the supersymmetric theory~\cite{Benvenuti:2021com}. These might still be stable in the presence of small SUSY breaking.

Since much of the foundational work has already been addressed in~\cite{Draper:2018mpj}, we conclude our discussion on the CP domain wall by presenting exact results for a specific case and estimating domain wall tension in several limiting scenarios.

\subsection{$F=1, N=2$: Exact Results Again}
For $SU(2)$, it is possible to derive the profile and tension of the domain wall, which connects the two degenerate CP breaking vacua from \Eq{eq:su2df}.
Identifying $\eta' = 2\delta_f$, the equation of motion that describes the field configuration along the $z$-axis perpendicular to the wall is given by
\beq
f^2 \frac{\dd^2 \eta'}{\dd z^2} = \frac{2|\Lambda|^{5/2}}{\sqrt x}[m(m_Q - 5x) - 8m_Q x \cos(\eta')]\sin(\eta').
\eeq
This differential equation is analytically solvable and the domain wall boundary condition requires that $\eta'$ approaches the two VEVs as $z\to\pm\infty$:
\beq
\lim_{z \to \pm\infty} \eta'(z) \rightarrow \pm \cos^{-1} \left( \frac{m(m_Q - 5x)}{8m_Q x} \right),
\eeq
which in turn singles out a desired wavefunction:
\beq
\eta'(z) = 2 \tan^{-1} \left[ \frac{\tanh\left(\frac{1}{4}z \sqrt{\frac{\Lambda^{5/2}(64m_Q^2 x^2 - m^2 (m_Q - 5x)^2)}{f^2 m_Q x^{3/2}} } \right)}{\sqrt{\frac{8m_Q x + m m_Q - 5mx}{8m_Q x - m m_Q + 5 m x}}} \right].
\label{eq:Vell}
\eeq
Since we have an analytical $\eta'$ field configuration, the domain wall tension can be easily computed exactly
\beq
T_{\eta'} = |\Lambda|^{5/2} \sqrt{ \frac{64m_Q^2 - m^2 (m_Q f^4 / |\Lambda|^5 - 5)^2}{m_Q} }.
\label{eq:dwTsu2}
\eeq
This tension decreases with decreasing $m_Q$ until the critical point $m_{Q,0}$ in \Eq{eq:su2mqmk0} is reached from above. For even lower masses there is only one stable minimum and no domain walls exist. This can be seen from the tension becoming imaginary when $m_Q$ is below the critical point. Note also that in the supersymmetric limit $m\rightarrow 0$, the tension $T=8 m_Q^{1/2} |\Lambda|^{5/2}$ agrees exactly with the tension of the BPS saturated domain-wall~\cite{Dvali:1996xe,Kovner:1997ca,Witten:1997ep,Chibisov:1997rc}.

\subsection{Limiting Cases}
Beyond $F=1$, $N=2$ finding analytic solutions for the domain wall profile becomes infeasible.
However, we can still study the CP domain walls in the limits of large $N$, small $m$ and $m_Q$, along possible domain wall trajectories: $\eta'$, pure pion and pion-like (mixture of $\eta'$ and pion) directions. Withtout loss of generality, we pick $\pi^{F-1}$ for the pure pion and $\phi \equiv \eta' - \pi^{F-1}$ for the pion-like paths. 
Again it is assumed that all quark masses are equal. Note that not all of these trajectories exist for each choice of $F$ and $N$. The estimates for the tension apply when stable domain walls exist in the corresponding field directions.

Let us first work in the large $N$ scenario, assuming that $m_Q^2/m^2 \gg 1/N$, i.e. the $\eta'$ becomes massless in the $m_Q\rightarrow 0$ limit as the contributions to the mass of the order $m^2/N$ are always smaller. In this limit the minima are at $\delta_f =\mathcal{O}(1/N)$ and $\delta_f = 2\pi - \mathcal{O}(1/N)$ (see Eq.~\eqref{eq:x_N}). Thus the minima would be close in the $\eta'$ direction if $\eta'$ were trivially $2\pi$-periodic. However, due to the existence of the branches and the corresponding non-differentiable points at their crossing, it is more favorable to traverse $\Delta\eta' \sim 2\pi$ if the minima can be connected in this direction. Note that domain walls in the $\eta' - \pi^{F-1}$ direction, in contrast, connect field values with a separation of the order $2\pi/F$.

The relevant terms in the Largrangian for the $\eta'$ direction in the large $N$ limit read 
\beq
\mathcal L_{\eta'}|_{N\gg1} \simeq \frac12 F f^2  (\partial \eta')^2 + 2 F m_Q ( f^2 m + 2|\Lambda|^3 ) \cos(\eta'),
\eeq
where the value of $f$ (or $x$) at the minimum can be found in \Eq{eq:x_N}. Given a Lagrangian, one can find the domain wall tension by solving the equation of motion for $\eta'$ that interpolates between the two degenerate VEVs and plugging it into
\beq
T_{\eta'}|_{N\gg1} = \int \dd z \left( \frac12 F f^2  \left(\frac{d \eta'}{d z}\right)^2 - 2 F m_Q ( f^2 m + 2|\Lambda|^3 ) \cos(\eta') \right).
\eeq
Further, we can estimate the domain wall tension by rescaling the coordinate $z$ and $\eta'$ such that the field value at the minima is $\mathcal{O}(1)$
\beq
T_{\eta'}|_{N\gg1} = F f \sqrt{m_Q(f^2 m + 2|\Lambda|^3)} \int \dd z \left( \frac12 \left(\frac{d \eta'}{d z}\right)^2 -  \cos(\eta') \right).
\eeq
Assuming that the integral over the dimensionless field gives an $\mathcal{O}(1)$ number, the domain wall tension in the $\eta'$-direction is then estimated as
\beq
T_{\eta'}|_{N\gg1} \sim F f \sqrt{|\Lambda|^3 m_Q},
\label{eq:tetalN}
\eeq
where we used that $f^2 m \lesssim |\Lambda|^3$ as can be seen from~\Eq{eq:x_N}. The domain wall profile is approximately, i.e. to leading order in $1/N$, given by
\beq
\eta'(z)|_{N\gg1} = 4 \tan^{-1} (e^{\tilde m z})
\eeq
with
\beq
\tilde m^2 \equiv 2m_Q \left( m + 2|\Lambda|^3/f^2 \right).
\eeq

Let us now move on to the other directions. 
The Lagrangian for the pion-like trajectory is given by 
\beq
\mathcal L_{\phi} \simeq \frac18 F^2 f^2  (\partial \phi)^2 +   2 m_Q ( f^2 m + 2|\Lambda|^3 ) \cos\left(\frac{F}{2}\phi\right),
\eeq
which for a field value separation of $\Delta \phi \sim 2\pi/F$ (see Eq.~\eqref{eq:dgspe}) yields the following scaling of the tension
\beq
T_{\phi}|_{N\gg1} \sim f \sqrt{|\Lambda|^3 m_Q}
\label{eq:tphilN}
\eeq
and profile
\beq
\phi(z)|_{N\gg1} = \frac8F \tan^{-1} (e^{\tilde {m} z}).
\eeq
\begin{table}[t]
\centering
\renewcommand{\arraystretch}{1.5}
\begin{tabular}{c|c|c|c}
\hline
                                                   & $F<N$                 & $F=N$                 & $N<F<3N/2$           \\ \hline
$T_{\eta'} |_{m\rightarrow0}$                      & $F |\Lambda|^{\frac{3N-F}{N}} m_Q^{\frac{F}{N}}$ & $N^{-\frac12} |\Lambda|^3$, $m_Q\Lambda^2$ & $F |\Lambda|^{\frac{5N-F}{2N}} m_Q^{\frac{N+F}{2N}}$ \\ \hline
$T_{\pi^{F-1}} |_{m\rightarrow0}$                      & $-$ & $m_Q |\Lambda|^2$ & $-$ \\ \hline
$T_{\phi} |_{m\rightarrow0}$                      & $|\Lambda|^{\frac{3N-F}{N}} m_Q^{\frac{F}{N}}$ & $-$
& $|\Lambda|^{\frac{5N-F}{2N}} m_Q^{\frac{N+F}{2N}}$ \\ \hline
$T_{\eta'} |_{m_Q\rightarrow0}$                      & $N^{-\frac{1}{2}}F^{\frac{3}{2}} |\Lambda|^{\frac{3N-F}{N}} m^{\frac{F}{N}}$ & $N^{-\frac12} |\Lambda|^3$ & $F |\Lambda|^{\frac{6N-4F}{2N-F}} m^{\frac{F}{2N-F}}$ \\ \hline
\multicolumn{1}{l|}{$T_{\phi} |_{m_Q\rightarrow0}$} & $N^{-\frac{1}{2}} |\Lambda|^{\frac{3N-F}{N}} m^{\frac{F}{N}}$ & $-$
& $F^\frac{3}{2} |\Lambda|^{\frac{6N-4F}{2N-F}} m^{\frac{F}{2N-F}}$ \\ \hline
\end{tabular}
\caption{CP domain wall tension scaling for different theories and domain wall trajectories. All quarks are taken to be degenerate and we only report the leading term in $F/N$ in the $F<N$ case. Note that the existence of all of these directions depends on $N$ and $F$.}
\label{tab:cpdwt}
\end{table}

In the same way, the scaling for the tension of the domain wall and its profiles can be found for $m\rightarrow0$ or $m_Q\rightarrow0$ for the potentials in Eqs~(\ref{eq:VellmQ}), (\ref{eq:VFgN}), and (\ref{eq:VFeqN}). We summarize the scaling of the $\eta'$, $\pi^{F-1}$ and $\phi$ domain wall tensions for $F<N$, $F=N$, and $N<F<3N/2$ in Table~\ref{tab:cpdwt}.

Before closing, let us comment on a few of these results. If we take $N \rightarrow \infty$, we get $T \sim |\Lambda|^3$ (similar to pure Yang-Mills) for $F < N$. In addition, when $F=N$ and $N$ is small, $\eta'$ is heavy, so the domain wall tension associated with it is also much larger than that for $\pi^{F-1}$, implying the $\eta'$ and $\phi$ directions are unstable and pure pionic domain walls are preferred. However, as $N$ increases, the $\eta'$ domain wall, if allowed, becomes the preferred direction again. 

%
\section{Conclusion}\label{sec:conclusion}
%
We have investigated the phase structure of a QCD-like theory
at $\bar\theta = \pi$ and found the phase diagram for spontaneous CP breaking.
The theory under discussion is the  $\mathcal N=1$ supersymmetric version of QCD with supersymmetry breaking added via anomaly mediation (AMSB).  The massless degrees of freedom in the far UV match those in ordinary QCD, as does the global symmetry breaking pattern in the IR.  Hence it is possible to compute the potential for the $\eta'$  and the neutral GBs as well as the entire the chiral Lagrangian for small SUSY breaking $m\ll\Lambda $.

Carefully examining the potential with a general spectrum of quark masses and SUSY breaking scale, we observe the emergence of a critical quark mass threshold $m_{Q,0} \propto m/N$. At this critical quark mass the  $\eta'$ becomes massless, and the CP symmetry is spontaneously broken for $m_Q > m_{Q,0}$, independent of the number of colors $N$. For $F<N$, the critical mass $m_{Q,0}$ for the lightest quark agrees with the $F=1$ scenario as long as the other quark masses far exceed $m$, and the massless mode at this transition corresponds to a mixture of $\eta'$ and pions. When $F>1$ with uniform quark masses, CP spontaneously breaks, but remains conserved if we have at least one massless quark. 
Similar phenomena are observed for $N+1<F<3N/2$, analogous to $F<N$, as well as for $F=N, N+1$ theories. 

We have also shown that an exact solution for the CP domain wall between two vacua can be found for $SU(2)$ and approximate solutions are available in special limiting cases for all theories investigated in this work.

\section*{Acknowledgements}
We would like to thank Clay Cordova, Sungwoo Hong, Zohar Komargodsky and Shimon Yankielowicz for useful discussions and feedback. 
The authors are supported in part by the NSF grant PHY-2014071. MR is also supported by a Feodor–Lynen Research Fellowship awarded by the Humboldt Foundation. CC is also funded in part by the US-Israeli BSF grant 2016153.
TY is also supported by the Samsung Science and Technology Foundation under Project Number SSTF-BA2201-06.

\appendix

\section{Potentials and Minima for $F=1$}
\label{sec:potandmin}
In this appendix, we describe how exact solutions are found for the $F=1$ potential. 
Given the potential in \Eq{eq:vlf1}, the minimization conditions take the form
\beq
0\stackrel{!}{=} \frac{\partial V}{\partial x}, \qquad
0\stackrel{!}{=} \frac{\partial V}{\partial \delta_f}.
\eeq
This can be solved analytically for $N=2,3$ as we will show in the following.

\subsection*{$SU(2)$ Solutions}
For $SU(2)$, the potential and minimization conditions simplify to
\begin{equation}
	V_{SU(2)} = \frac{2 \Lambda^{5/2}}{\sqrt{x}} \left(m_Q^2 + x^2 - m (m_Q - 5 x) \cos\left(2 \delta_f\right) + 
   2 m_Q x \cos\left(4 \delta_f\right)\right)
   \label{eq:vsu2}
\end{equation}
and
\begin{align}
	0&= -m_Q^2 + 3 x^2 + m (m_Q + 5 x) \cos(2 \delta_f) +  2 m_Q x \cos(4 \delta_f)\,,\\
	0&= \left(m (m_Q - 5 x) - 8 m_Q x \cos(2 \delta_f)\right) \sin(2 \delta_f)\,,
\end{align}
The last equation has two sets of solutions
\begin{enumerate}
	\item $\sin(2\delta_f)=0$, i.e. $\delta_f =0,\pi/2$ for $\delta_f\in [0,\pi)$. Plugging these two values into the minimization equation for $x$ one obtains a quadratic equation for $x$. Altogether one obtains four solutions
		\begin{align}
			x_{1/2} &= -\frac{1}{6} \left(5 m + 2 m_Q \pm \sqrt{25 m^2 + 8 m m_Q + 16 m_Q^2}\right)\,,& &\delta_f = 0\,,\\
			x_{3/4}&= \frac{1}{6} \left(5 m - 2 m_Q + \sqrt{25 m^2 - 8 m m_Q + 16 m_Q^2}\right)\,, & &\delta_f = \frac{\pi}{2}\,.
		\end{align}
	\item $m (m_Q - 5 x) - 8 m_Q x \cos(2 \delta_f) = 0$. This allows us to solve for $\delta_f$
		\begin{equation}
			\delta_f = \pm \frac{1}{2}\arccos \left( \frac{m (m_Q - 5 x)}{8 m_Q x} \right)\,.
                \label{eq:su2df}
		\end{equation}
		Plugging this into the minimization equation for $x$ one obtains a cubic equation for $x$
		\begin{equation}
			\frac{3 m^2 m_Q}{16} - \left(\frac{5 m^2}{8} + m_Q^2\right) x - \left(\frac{25 m^2}{16 m_Q} - 2 m_Q\right) x^2 +   3 x^3 = 0\,.
		\end{equation}
		This has three real roots (the discriminant is positive) which can be expressed as
		\begin{equation}
			x_k =2 \sqrt{-\frac{p}{3}} \cos \left(\frac{1}{3} \arccos \left(\frac{3 q}{2 p} \sqrt{\frac{-3}{p}}\right)-k \frac{2 \pi}{3}\right) + b_0\quad \text{for}\quad k=0,1,2.
		\end{equation}
		with
		\begin{align}
			p&= -\frac{3040 m^2 + \tfrac{625 m^4}{m_Q^2} + 3328 m_Q^2}{6912}\,,\\
			q&=-\frac{15625 m^6 + 114000 m^4 m_Q^2 + 139008 m^2 m_Q^4 + 143360 m_Q^6}{1492992 m_Q^3}\,,\\
			b_0 &= \frac{25 m^2}{144 m_Q} + \frac{2 m_Q}{9}\,.
		\end{align}
		Thus one obtains six further solutions. Among the three $k$'s, $k=0$ yields the global minimum. 
\end{enumerate}
The critical value can be found analytically by requiring that $\partial^2 V/\partial\delta_f^2 =0$ at the minimum, i.e.
\begin{equation}
	0 = \frac{\partial^2 V}{\partial\delta_f^2} \propto m (m_Q -5x) \cos(2\delta_f ) - 8m_Q x \cos(4\delta_f)\,.
\end{equation}
We know that for $m_Q = 0$ the solution is $\delta_f = \tfrac{\pi}{2}$ and
\beq
x_{3/4}= \frac{1}{6} \left(5 m - 2 m_Q \pm \sqrt{25 m^2 - 8 m m_Q + 16 m_Q^2}\right).
\eeq
Plugging this into the equation we find
\begin{equation}
	24 m_Q (25m^3 -54 m^2 m_Q + 32 m_Q^3) = 0\,,
\end{equation}
which is a cubic equation for $m_Q$ with three real solutions
\begin{equation}
	m_{Q,0} = \frac{3m}{2} \cos\left( \frac{1}{3} \arccos \left(-\frac{25}{27}\right) -\frac{2\pi k}{3} \right)\,,\quad \text{with}\quad k=0,1,2\,.
 \label{eq:su2mqm}
\end{equation}
The $k=1$ solution agrees with the numeric determination of the critical point and yields $m_Q/m \approx 0.576511$.

\subsection*{$SU(3)$ Solutions}
For $SU(3)$ the potential is of the form
\begin{equation}
	V_{SU(3)} = \frac{2 \Lambda^{8/3}}{x^{2/3}} \left(m_Q^2 + x^2 - m m_Q \cos(2 \delta_f) -  8 m x \sin(\ell \pi + \delta_f) -   2 m_Q x \sin (\ell \pi + 3 \delta_f)\right)
\end{equation}
with the minimization conditions
\begin{align}
	0&= m_Q^2 - 2 x^2 - m m_Q \cos(2 \delta_f ) + 
 4 m x \sin (\ell \pi + \delta_f ) + m_Q x \sin (\ell \pi + 3 \delta_f )\,,\\
	0&= -8 m x \cos (\ell \pi  + \delta_f ) - 6 m_Q x \cos (\ell \pi + 3 \delta_f) +  2 m m_Q \sin (2 \delta_f)\,.
\end{align}
A major difference compared to the $SU(2)$ scenario is that here we have a branched potential with $\ell=0,1$. The actual potential will be the lower envelop of the two branches. The branches are required to restore the $2\pi$ periodicity of $\eta'$.

Note that the $\ell =0$ and $\ell=1$ equations are related by a shift of $\delta_f \rightarrow\delta_f +\pi$. Thus we can solve the minimization conditions for $\ell = 0$ and obtain the $\ell=1$ solutions by making the above shift.

For $\ell=0$ the minimization equations simplify to
\begin{align}
	0&= m_Q^2 - 2 x^2 - m m_Q \cos(2 \delta_f ) + 
 4 m x \sin (\delta_f ) + m_Q x \sin ( 3 \delta_f )\,,\\
	0&= \cos(\delta_f ) \left(4 m x - 3 m_Q x + 6 m_Q x \cos (2 \delta_f ) - 
   2 m m_Q \sin(\delta_f)\right)\,.
\end{align}
The equation for the phase is solved by
\begin{enumerate}
	\item $\cos(\delta_f )=0$, i.e. $\delta_f = \pm \pi/2$. For these values the equation for $x$ becomes quadratic with the solutions
	\begin{align}
		x_{1/2} &=\frac{1}{4}\left(4m - m_Q \pm \sqrt{16m^2+9 m_Q^2}\right)\,,\quad &\delta_f = \frac{\pi}{2}\,,\\
		x_{3/4} &=\frac{1}{4}\left(m_Q - 4m \pm \sqrt{16m^2+ 9 m_Q^2}\right)\,,\quad &\delta_f = -\frac{\pi}{2}\,.
	\end{align}
	\item $(4 m  - 3 m_Q + 6 m_Q \cos (2 \delta_f )) x - 2 m m_Q \sin(\delta_f)=0$. This still possesses analytic solutions but this requires solving a quartic equation, leading to long and horrible expressions.
\end{enumerate}
The critical point is determined by also requiring that
\begin{equation}
	0= \frac{\partial^2 V}{\partial \delta_f^2}\propto 4 m m_Q \cos (2 \delta_f ) + 8 m x \sin (\delta_f) +  18 m_Q x \sin (3 \delta_f)\,.
\end{equation}
The $\delta_f = \pi/2$ minimum ends at the critical point. Thus we plug in $\delta_f = \pi /2$ and $x_{1/2} =\frac{1}{4}\left(4m - m_Q \pm \sqrt{16m^2+9 m_Q^2}\right)$, what yields
\begin{equation}
	0 = -6m_Q \left(16m^3 -48 m^2 m_Q +9 m m_Q^2 +27 m_Q^3\right)\,.
\end{equation}
This is solved for $m_Q = 0$ and
\begin{equation}
	m_{Q,0} = \frac{14m}{9} \cos\left( \frac{1}{3} \arccos \left(-\frac{289}{343}\right) -\frac{2\pi k}{3} \right)- \frac{m}{9}\,,\quad \text{with}\quad k=0,1,2\,.
 \label{eq:su3mqm}
\end{equation}
The $k=1$ solution yields $m_{Q,0} = 0.398853$ which is what we also find numerically.

\subsection*{Large $N$}
To leading order in the large $N$ limit for one flavor it is straightforward to obtain an analytic expression for the critical point $m_{Q,0}$. We make a perturbative ansatz, i.e.
\begin{equation}
x = x_0 + x_{-1}/N\,,\quad \eta' = \eta'_0 + \eta'_{-1}/N\,,\quad m_{Q,0} = m_{Q,0_0} + m_{Q,0_{-1}}/N\,,
\end{equation}
where we have made the identification $\eta' = 2\delta_f$. Using the defining equations for the critical point
\beq
\frac{d V_\ell}{d x} =0\,,\quad \frac{d V_\ell}{d \eta'} = 0\,,\quad \frac{d^2 V_\ell}{d\eta'^2}=0\,,
\eeq
we find to leading order in $1/N$ the following equations
\begin{align}
-m_{Q,0_0}^2+ x_0(x_0-3m)+m m_{Q,0_0} \cos (\eta'_0) &= 0\,,\\
m_{Q,0_0} (m + 2 x_0) \sin (\eta'_0) &=0\,,\\
m_{Q,0_0}(m+2 x_0) \cos (\eta'_0) &=0\,.
\end{align}
For positive $x_0$ this can only be solved by $x_0 = 3m$ and $m_{Q,0_0} = 0$ with $\eta'_0$ undetermined. Plugging this into the equations for the next to leading order in $1/N$ we obtain
\begin{align}
3(4m + x_{-1}) + m_{Q,0_{-1}} \cos (\eta'_0 ) &= 0\,,\\
9m(2\pi \ell + \eta'_0 - \theta) + 7 m_{Q,0_{-1}} \sin (\eta'_0) &=0\,,\\
9m + 7 m_{Q,0_{-1}} \cos(\eta'_0 ) &=0\,. 
\end{align}
The first and third equation immediately give $x_{-1} = -\tfrac{25}{7}m$. The second and third equation in general do not have a solution. However, for $\theta = \pi$ they are solved by $\eta'_0 = \pi - 2\pi \ell$ and $m_{Q,0_{-1}} = \tfrac{9}{7}m$, s.t. we find in summary
\begin{equation}
x = 3m - \frac{25}{7} \frac{m}{N}\,,\quad \eta' = \pi - 2\pi \ell\,,\quad m_{Q,0} = \frac{9}{7}\frac{m}{N}\,.
\label{eq:lNmqm}
\end{equation}
%

\bibliographystyle{JHEP}
\bibliography{bibtex}{}
\end{document}